\documentclass[reqno,12pt]{article}
\usepackage{ae} 
\usepackage[T1]{fontenc}
\usepackage[ansinew]{inputenc}
\usepackage{amsmath}
\usepackage{amssymb}
\usepackage{graphicx}
\usepackage{color}
\usepackage[colorlinks]{hyperref}
\usepackage{epsfig}

\newcommand{\beq}{\begin{equation}}
\newcommand{\eeq}{\end{equation}}
\newcommand\beqa{\begin{eqnarray}}
\newcommand\eeqa{\end{eqnarray}}
\newcommand\bea{\begin{array}}
\newcommand\eea{\end{array}}

\def\XXint#1#2#3{{\setbox0=\hbox{$#1{#2#3}{\int}$}
\vcenter{\hbox{$#2#3$}}\kern-.5\wd0}}

\newcommand{\nn}{\nonumber}

\newcommand{\neqa}{\nonumber\end{eqnarray}}
\newcommand{\la}[1]{\label{#1}}

\newcommand{\rf}[1]{(\ref{#1})}

\newcommand{\Tr}{{\rm Tr}}

\renewcommand{\d}{\partial}

\newcommand{\<}{{\langle}}
\renewcommand{\>}{{\rangle}}

\newcommand{\cA}{{\cal A}}

\newcommand{\cL}{{\cal L}}

\newcommand{\re}{\relax{\rm I\kern-.18em R}}

\newcommand{\sh}{{\rm sinh}}
\newcommand{\ch}{{\rm cosh}}

\renewcommand{\sp}{p\hspace{-.40em}/}

\def\su2{{SU(2)}}

\def\[{\left[}
\def\]{\right]}

\def\({\left(}
\def\){\right)}
\def\[{\left[}
\def\]{\right]}

\def\<{\langle}
\def\>{\rangle}

\def\i2{\frac{i}{2}}

\def\bS{{\mathbb S}}

\def\cJ{{\cal J}}
\def\spi{\relax{\rm \pi\kern-0.5em /}}
\def\sA{\relax{\rm A\kern-0.5em /}}
\def\sp{\relax{\rm p\kern-0.5em /}}
\def\sd{\relax{\rm \d\kern-0.5em /}}
\def\sk{\relax{\rm k\kern-0.5em /}}
\def\sn{\relax{\rm n\kern-0.5em /}}
\def\sl{\relax{\rm l\kern-0.5em /}}
\def\sP{\relax{\rm P\kern-0.7em /}}
\def\sbeta{\relax{\rm \beta\kern-0.5em /}}
\def\cN{{\cal N}}

\usepackage{varioref}
\usepackage{makeidx}
\makeindex

\usepackage[english]{babel}
\begin{document}


\thispagestyle{empty}
\begin{flushright}\footnotesize
\texttt{AEI-2009-078}\\
\vspace{1.7cm}
\end{flushright}

\renewcommand{\thefootnote}{\fnsymbol{footnote}}
\setcounter{footnote}{0}
\setcounter{figure}{0}
\begin{center}
{\Large\textbf{\mathversion{bold} Symmetries of the $\cN=4$ SYM S-matrix }\par}


\vspace{1.0cm}

\textrm{Amit Sever$^a$ and Pedro Vieira$^{\dot a}$}
\\ \vspace{1.2cm}
\footnotesize{
\textit{$^a$
Perimeter Institute for Theoretical Physics\\ Waterloo,
Ontario N2J 2W9, Canada} \\
\texttt{asever AT perimeterinstitute.ca}
\vspace{3mm}

\textit{$^{\dot a}$ Max-Planck-Institut f\"ur Gravitationphysik, Albert-Einstein-Institut, \\ Am M\"uhlenberg 1, 14476 Potsdam, Germany;  } \\
\texttt{pedrogvieira AT gmail.com}
\vspace{3mm}}


\par\vspace{1.5cm}

\textbf{Abstract}\vspace{2mm}
\end{center}

\noindent
\small

Under the assumption of a CSW generalization to loop amplitudes in $\cN=4$ SYM,
\begin{enumerate}
\vspace{-.35cm}
\item We prove that, formally the S-matrix is superconformal invariant to any loop order, and
\vspace{-.2cm}
\item We argue that superconformal symmetry survives regularization. More precisely, IR safe quantities constructed from the S-matrix are superconformal covariant. The IR divergences are regularized in a new holomorphic anomaly friendly regularization.
\end{enumerate}
\vspace{-.35cm}
The CSW prescription is known to be valid for all tree level amplitudes and for one loop MHV amplitudes. In these cases, our formal results do not rely on any assumptions.

\vspace*{\fill}

\setcounter{page}{1}
\renewcommand{\thefootnote}{\arabic{footnote}}
\setcounter{footnote}{0}

\newpage


{\footnotesize
\tableofcontents
}

\newpage

\section{Introduction and Discussion}

Symmetry has proved to be of utmost importance in unveiling the remarkable beauty hidden in $\mathcal{N}=4$ super Yang-Mills. Two examples illustrate this rather nicely:
\begin{enumerate}
\item The study of the planar spectrum of this gauge theory is mapped to the study of an integrable model \cite{first}. Particle excitations in this model transform under an extended $SU(2|2)$ symmetry algebra which completely constrains the $2$-body S-matrix \cite{second}, the main ingredient in the computation of the exact spectrum of the theory.\footnote{In an integrable theory, finding the S-matrix is the main step towards the computation of the exact spectrum which follows a (not completely) standard recipe \cite{Zamolodchikov:1989cf}, carried out in the AdS/CFT context in \cite{first,second,BES,third} and references therein.}  

\item A second example, closely related to the subject of this paper, concerns planar scattering amplitudes in $\mathcal{N}=4$ SYM. Both at weak and strong coupling, these amplitudes possess an anomalous {\it dual conformal symmetry} \cite{dualpapers,dualpapers2,dualpapers3,AM, BM}. For $4$ and $5$ particles, this anomalous symmetry fixes the form of the Maximally Helicity Violating (MHV) amplitudes at any value of the t' Hooft coupling in terms of the cusp anomalous dimension \cite{dualpapers2} which can be computed from Integrability \cite{BES}. For more than $5$ particles, this symmetry is still very constraining. It fixes the form of the MHV planar amplitudes to be given by the BDS ansatz \cite{BDS} times a reminder function that depend only on the dual conformal cross-ratios and become trivial in collinear limits. Surprisingly, it is the mysterious dual superconformal symmetry that is under control at loop level whereas the usual superconformal symmetry is understood at tree level only \cite{tree1,tree2,Bargheer:2009qu,Korchemsky:2009hm}.  
\end{enumerate}

There seems to be two deep connections between these two points. First, the usual conformal symmetry as well as the dual conformal symmetry of $\mathcal{N}=4$ SYM form a Yangian \cite{Ystrong1,BM,Ystrong2,tree2} -- the structure of higher charges arising in integrable models. Second, as emphasized in \cite{Bargheer:2009qu}, the superconformal generators which act on the generating function of scattering amplitudes, are expected to share many features with the length changing higher charges that acts on a single trace operators. These appear in the context of computing the planar spectrum of the theory. The possibility of making such nice connections precise in the future, as well as the remarkable success observed so far, entitles us to big expectations.

At tree level, partial scattering amplitudes were shown to be invariant under superconformal transformations \cite{Witten,tree2}. However, due to the so called holomorphic anomaly \cite{CSW}, superconformal transformations fail to annihilate the tree level amplitudes whenever two adjacent momenta become collinear. As was shown in \cite{Bargheer:2009qu}, that failure can be corrected by adding a term to the superconformal generators that split one massless particle into two collinear ones. As we will show in this paper, already at tree level, there are additional points in phase space where superconformal transformations fail to annihilate a tree level amplitude. E.g.,  the points where the amplitude factorize on a multi-particle pole and, in addition, the on-shell internal momenta become collinear to one of the external momenta.  At these points, the failure can be corrected by adding a term to the superconformal generators that at tree level, join two disconnected amplitude with a shared collinear momenta into a single connected amplitude. The resulting symmetry is therefore not a symmetry of an individual amplitude but instead, a symmetry of the tree level S-matrix. That correction of the tree level generators might seem like a picky detail -- after all, for generic momenta, the tree level amplitudes are indeed symmetric. However, at loop level this detail becomes of major importance. The reason is that internal loop momenta scan over all phase space and in particular on the points were they become collinear to an external momenta. As a result, the superconformal generators fails to annihilate {\it any} loop amplitude. That is clearly \textit{not} a picky detail!

Further complication of loop amplitudes over tree level ones is the presence of IR divergences and the resulting need for regularization. These IR divergences arise from the region of integration where an internal momenta become collinear to an external one. Therefore, the IR divergences and the failure of superconformal invariance are closely related.

The goal of this paper is to promote the superconformal symmetry to loop level. We will assume that the MHV diagrammatic expansion \cite{CSW} holds at any loop order, which although very plausible was only checked in the literature to one loop \cite{Brandhuber:2004yw}. Under that assumption, we will find a correction to the generators that annihilate the full S-matrix.

The analysis will first be done without a regulator. Such analysis is only formal because $\cN=4$ YM is conformal and therefore don't have asymptotic particles. As a result, the probability for scattering some fixed number of massless particles into another fixed number of massless particles is zero. Technically, the corresponding perturbative analysis is plagued with IR divergences. The S-matrix is however non-trivial by means that there are physical observables constructed from the would have been S-matrix. These are IR safe quantities such as inclusive cross sections\footnote{These usually involve an external probe. See \cite{Bork:2009ps} for a recent study of these in $\cN=4$ SYM.}. In perturbation theory, the only known way to construct these observables is from the S-matrix elements which, for massless particles, are not good observables.\footnote{Ideally, one would like to have an alternative formulation of physical IR safe observables that do not pass through the IR unsafe S-matrix. Identifying these observables in the T-dual variables \cite{AM} may help in finding such formulation.} To overcome that problem, one first introduces an IR regulator. The resulting IR regulated theory has an S-matrix from which the desired observables are computed. A good IR regulator is a regulator that drops out of IR safe physical quantities leaving a consistent answer behind. We will argue that these physical observables are superconformal covariant. That is, we will show that no violation of superconformal invariance emerges from their (regulated) S-matrix elements building blocks.

A similar issue arises in the study of dual conformal invariance of planar amplitudes. In analogy to conformal symmetry, there, loop amplitudes are formally dual conformal covariant.  However, any regularization result in a dual conformal anomaly \cite{dualpapers2,Elvang:2009ya}. The anomaly however, can be recast as a correction to the dual conformal generators such that they act on the regulator.\footnote{As far as we know, that point was not illustrated in the literature. We have checked that explicitly in two different regularizations. One, is the regularization in which the external particles are given a small mass. The other, is the Alday-Maldacena regularization \cite{AM} where scattered particles are charged under a small gauge group on the Coulomb branch of the large $N$ one. Note added to this footnote: As this paper was being completed the work \cite{Alday:2009zm} appeared where this scenario was checked at one loop in the Coulomb Branch regularization, see \cite{Alday:2008yw} for a related discussion at strong coupling.} In other words, the corrected generators are anomaly free. Planar IR safe quantities are therefore dual superconformal covariant as no violation of dual conformal invariance arises from their S-matrix elements building blocks.

The situation with conformal transformations is more involved. The reason is that the terms in the superconformal generators that, at tree level, join two disconnected amplitudes can also act on a connected part of an amplitude. When they do, a new loop is formed within the connected amplitude. It is therefore not enough for regulate the amplitudes but we must also regulate the superconformal generators.

In the last section, we will regulate the amplitudes and repeat the calculation, now identifying a regularized form of the generators. The calculation will be done in a new regularization in which the external particles are given a small mass. The regulated amplitude is then computed using the CSW prescription, now treating the external particle in the same way as the internal ones. We expect that planar dual-conformal invariance can be proved to all orders in perturbation theory using the same techniques.

The paper is organized as follows: In section \ref{sec:tree} we review the superconformal invariance of tree level amplitudes \cite{tree2,Bargheer:2009qu}. In section \ref{sec:fincuts} we find the unregularized form of the generators by demanding that they annihilate finite unitarity cuts of the MHV one loop generating function. In section \ref{sec:formal} we show that the generators found in section \ref{sec:fincuts} formally annihilate the unregularized one loop MHV generating function. In section \ref{all} we use the tree level results and a conjectured loop level CSW generalization to formally derive the superconformal invariance of the full S-matrix at any loop order. In section \ref{sec:reg} we will propose a new "holomorphic anomaly friendly" regularization named {\it sub MHV regularization}. It will allow us to keep the symmetry, found in the previous sections, under control and read of a regularized form of the generators. Appendix \ref{AppA} contains some technical details relevant to section \ref{sec:formal}.

\section{Superconformal Invariance of Tree Level Amplitudes}\la{sec:tree}

This section is a quick review of partial tree level partial scattering amplitudes in $\cN=4$ SYM, their generating function and superconformal invariance. We assume the reader is familiar with the spinor helicity formalism and only set notation and highlight a few essential points.

On-shell states in $\cN=4$ are most conveniently represented in spinor helicity superspace. The light-like momenta is decomposed into a product of a positive chirality spinor $\lambda^a$ and a negative chirality spinor $\widetilde\lambda^{\dot a}$ through $k^{a\dot a}=\lambda_a\widetilde\lambda_{\dot a}=s\lambda^a\bar\lambda^{\dot a}$. Here $\bar\lambda=(\lambda)^*=s\widetilde\lambda$ and $s$ is the energy sign of $k$. We will work in $(+,-,-,-)$ signature so that $s={\rm sign}(k^0)={\rm sign}(k_0)$. In superspace, the scattering amplitude of $n$ particles is a function
$$
\cA_n(\lambda_1,\widetilde\lambda_1,\eta_1,\dots,\lambda_n,\widetilde\lambda_n,\eta_n)~,
$$
where in our convention all particles are out-going and $\eta^A$, $A=1,\dots,4$ is a superspace coordinate transforming in the fundamental representation of the $SU(4)$ R-symmetry. Amplitudes can be classified by their helicity charge
\beq
h(n,k)=2n-4k~. \la{helicity}
\eeq
Here the number $k$ counts the number of $\eta$'s. It ranges between 8 for MHV amplitudes and $4n-8$ for $\overline{\rm MHV}$ amplitudes. The MHV partial amplitudes take the simple form \cite{MHVpapers}
\beq\label{MHVgen}
\mathcal{A}^{(0)\, \rm MHV}_n(1,2,\dots,n)=\delta^4(P_n)A_n^{(0)\,\rm MHV}(1,2,\dots,n) =i\frac{\delta^8(Q_n)\delta^4(P_n)}{\<12\>\<23\>\dots\<n1\>} ~,
\eeq
where $P_n=\sum_{j=1}^n k_j$ and $Q_n=\sum_{j=1}^n \lambda_j^a \eta_j^B$. Here, for keep notations simple, we omitted a factor of $g^{n-2}(2\pi)^4$. The dependence on the YM coupling $g$ will be restored below when defining the generating functional whereas, the $(2\pi)^4$ factor is systematically  removed from vertices and propagators in our conventions. The superconformal generators that will be most relevant for us are the special conformal generator and its fermionic counterparts, the special conformal supercharges\footnote{We use Latin letters to denote the superconformal generators. Vive la r\'esistance.}
\beq\la{Gen}
S_{a A}=\sum_{i=1}^n {\d\over\d\lambda^a_i}{\d\over\d\eta_i^A}~,\qquad\bar S_{\dot a}^{A} =  \sum_{i=1}^n \eta_i^A{\d\over\d\widetilde\lambda^{\dot a}_i}~,\qquad K_{a\dot a} =\sum_{i=1}^n {\d\over\d\lambda^a_i}{\d\over\d\widetilde\lambda^{\dot a}_i}~.
\eeq
They satisfy the commutation relation
\beq
\{S_{aA},\bar S_{\dot a}^B\}=\delta_A^BK_{a\dot a}~. \la{SSK}
\eeq
Furthermore, the $S$ generator can be obtained from $\bar S$ by conjugation. For this reason, in the next sections, we will mostly focus on the special conformal supercharge $\bar S$.

The superconformal generators annihilate all tree level amplitudes provided the external momenta are generic \cite{Witten,tree2}. However, due to the presence of the so-called holomorphic anomaly \cite{CSW}
\beq\la{holomorphicAnomaly}
{\d\over\d\bar\lambda^{\dot a}} \frac{1}{\< \lambda,\mu\>} = \pi\bar\mu_{\dot a} \delta^2(\<\lambda,\mu\>)
\eeq
the action of the generators (\ref{Gen}) on the tree level amplitude results in extra terms supported at the points in phase space where two adjacent momenta are collinear. At these points, the generators (\ref{Gen}) fail to annihilate an individual tree level amplitude. Physically, the reason is that two on-shell collinear massless particles and a single particle carrying their momenta and quantum numbers are undistinguishable and are mixed by the generators (\ref{Gen}). Therefore, a more suitable object to act on is the generating function of all connected amplitudes whose ordered momenta forms the same polygon shape in momentum space.\footnote{That is, the same polygon Wilson loop dual in the sense of \cite{AM}.} For simplicity, one may consider instead the generating function of all connected tree level amplitudes \cite{Bargheer:2009qu}
\beq
\mathcal{A}^{(0)}_c[J]
= \sum_{n=4}^\infty{g^{n-2}\over n} \int  d^{4|4}\Lambda_1\, \dots d^{4|4}\Lambda_n\sum_{s_j=\pm} \Tr \( J(\Lambda_1^{s_1})\ldots J(\Lambda_n^{s_n}) \) \, \mathcal{A}_n^{(0)}(\Lambda_1^{s_1},\ldots,\Lambda_n^{s_n})~, \la{genfun}
\eeq
where $\Lambda^\pm=(\lambda,\pm\bar\lambda,\eta)$ parametrizes the null momenta and polarizations and the $(0)$ superscript stands for tree level. The sum over $s_j=\pm $ accounts for positive and negative energy particles. The $n$-particle partial amplitude is then given by
$$
\cA_n^{(0)}(\Lambda_1,\dots,\Lambda_n)=\left.{1\over N_c^n}\Tr\({\delta\over\delta J(\Lambda_n)}\dots{\delta\over\delta J(\Lambda_1)}\)\cA_c^{(0)}[J]\right|_{J=0}~.
$$
When acting on the generating function, the special conformal supercharges take the form
\beqa\la{bare}
({S}_{{1\to 1}})_{aA}&=&\sum_{s=\pm} \int d^{4|4}\Lambda\,  \Tr\[\partial_a\partial_A J(\Lambda^s)\, \check{J}(\Lambda^s)\]~,\\(\bar{{S}}_{{1\to 1}})_{\dot a}^A&=&-\sum_{s=\pm}s\int d^{4|4}\Lambda\, \eta^A  \,  \Tr\[\bar\partial_{\dot a}J(\Lambda^s) \check{J}(\Lambda^s)\] \,,\nn
\eeqa
where
$$\check{J}(\Lambda)=\frac{\delta}{\delta J(\Lambda)}~,\qquad s\bar\d_{\dot a}=s{\d\over\d\bar\lambda^{\dot a}}={\d\over\d\widetilde\lambda^{\dot a}}\,,
$$
$\partial_A=\partial/\partial \eta^A$ 
and the $(1\to1)$ subscript indicate that they preserve the number of external particles. In \cite{Bargheer:2009qu} it was shown that a corrected version of the generators do annihilate the generating function. That is
\beq\la{treecorr}
\({\bar S}_{1\to 1}+g{\bar S}_{1\to 2}\)\mathcal{A}^{(0)}_c[J]=0~,
\eeq
where ${\bar S}_{1\to 2}$ splits a particle into two collinear ones. For the special conformal supercharges, these are given by\footnote{Here written in a slight different way than in  \cite{Bargheer:2009qu}, using for example $\bar\lambda\eta'=\bar\lambda_1\eta_2-\bar\lambda_2\eta_1$. Moreover, the overall sign $s$ in $\bar S$ seems to have been overlooked in \cite{Bargheer:2009qu}.}
\beqa \la{treegen}
&&(\bar S_{1\to2})_{\dot a}^A=+2\pi^2{\sum_{s,s_1,s_2=\pm}}\!\!\!\!\!'~s\int d^{4|4}\Lambda d^4\eta'd\alpha\bar\lambda_{\dot a}{\eta'}^A\Tr\[\check J(\Lambda^s)\hat J(\Lambda_1^{s_1}) J(\Lambda_2^{s_2})\]~, \\
&&(S_{1\to 2})_{Aa}=-2\pi^2{\sum_{s,s_1,s_2=\pm}}\!\!\!\!\!'~\int d^{4|4}\Lambda d^4\eta'd\alpha\delta^{(4)}(\eta')\lambda_a\d'_A\Tr\[\check J(\Lambda^s)\hat J(\Lambda_1^{s_1}) J(\Lambda_2^{s_2})\]~,
\eeqa
where
$$
\hat J(\Lambda)={1\over 2\pi}\int_0^{2\pi} d\varphi e^{2\varphi i}J(e^{i\varphi}\Lambda)
$$
and
$$
{\sum_{s,s_1,s_2=\pm}}\!\!\!\!\!'~=\sum_{s,s_1,s_2=\pm} \delta_{0,(s-s_1)(s-s_2)}
$$
is a sum over the energy signs $s$, $s_1$ and $s_2$ such that $s\in\{s_1,s_2\}$. For $s_1=s_2$
\beq\label{collinearityportion}
\begin{array}{l}
\lambda_1=\lambda\sin\alpha\\
\lambda_2=\lambda\cos\alpha
\end{array}~, \qquad \begin{array}{l}
\eta_1=\eta\sin\alpha-\eta'\cos\alpha\\
\eta_2=\eta\cos\alpha+\eta'\sin\alpha
\end{array}~,\qquad\alpha\in[0,{\pi\over 2}]~.
\eeq
The other two cases where $s=s_1=-s_2$ and $s=s_2=-s_1$ are related to (\ref{collinearityportion}) by replacing $\sin(\alpha)\to\sh(\alpha)$ and $\sin(\alpha)\to\ch(\alpha)$ correspondently. Moreover, the corrected generators were shown to close the same superconformal algebra (see \cite{Bargheer:2009qu} for details).

At tree level, (\ref{treecorr}) was claimed to hold at {\it any} point in phase space \cite{Bargheer:2009qu}. As we will see in section 5, there are extra points in phase space where the holomorphic anomaly contributes. These are the points where the tree level amplitude factorizes on a multi-particle pole and an internal momentum is collinear to one of the neighboring momenta. Similarly to $\bar S_{1\to2}$ correcting $\bar S_{1\to1}$, these are accounted for by the inclusion of two new corrections $\bar S_{2\to1}$ and $\bar S_{3\to0}$. These however (at tree level) act on two or three disconnected tree level partial amplitudes, joining them into a single connected amplitude. Therefore, the object that is superconformal invariant is not the generating function of all connected partial amplitudes (\ref{genfun}), but instead the generating function of all partial amplitudes
$$
\bS^{\rm tree}[J]=\exp{\cA_c^{(0)}[J]}\,,
$$
connected and disconnected. That is the interacting part of the tree level S-matrix (see section 5 for more details). For example, the generator $\bar S_{2\to 1}$ reads
$$
\bar S_{2\to1}=2\pi^2\sum_{s=\pm}s\int d^{4|4}\Lambda d^4\eta'd\alpha\bar\lambda\eta'\Tr\[J(\Lambda^s)\check J(\Lambda_1^s)\check J(\Lambda_2^s)\]~,
$$
where $\Lambda_1$ and $\Lambda_2$ are given in (\ref{collinearityportion}). The corrected tree level symmetry is therefore
\beq\la{coeetreesymmetry}
\({\bar S}_{1\to 1}+g{\bar S}_{1\to 2}+g\bar S_{2\to1}+g\bar S_{3\to0}\)\bS^{\rm tree}[J]=0~.
\eeq
This structure generalizes to loop level
\beq\la{coeeloopsymmetry}
\({\bar S}_{1\to 1}+g{\bar S}_{1\to 2}+g\bar S_{2\to1}+g\bar S_{3\to0}\)\bS[J]=0~,
\eeq
where $\bS=\exp\cA_c[J]$ and
\beq
\mathcal{A}_c[J]= \sum_{n=4}^\infty \sum_{l=0}^\infty g^{2l}\mathcal{A}_n^{(l)}[J]
\eeq
is the connected generating function of scattering amplitudes. Here, $(l)$ stands for the number of loops and $\cA^{(l)}_n[J]$ is defined as in (\ref{genfun}).

Note in particular that the generators do not receive higher loop corrections. The full ``$\cN=4$ S-matrix" is obtained from $\bS$ by adding the forward amplitudes where some of the particles do not interact. The quotation marks are to remind the reader that, before regularization, $\cN=4$ SYM is conformal and therefore has no S-matrix. The correction of the formal relation (\ref{coeeloopsymmetry}) due to the IR regularization will be discussed in the last section.

The aim of this paper is to show that indeed the tree level symmetry (\ref{coeetreesymmetry}) generalizes to the loop level (\ref{coeeloopsymmetry}).

\section{Superconformal Invariance of One Loop Unitarity Cuts}\label{sec:fincuts}

Unitarity cuts of an amplitude are physical observable that compute the total cross section in the corresponding channels \cite{Cutkosky:1960sp,cut2}. These are always less divergent than the full loop amplitude and therefore can provide finite, regularization independent, information. In this section we will compute the finite cuts of $\bar S_{1\to 1} \cA$ at one loop and for $n$ final particles. By doing so, we will obtain an unregularized version of the generators and postpone the issue of regularization to latter sections.

\begin{figure}[t]
\epsfxsize=7cm
\centerline{\epsfbox{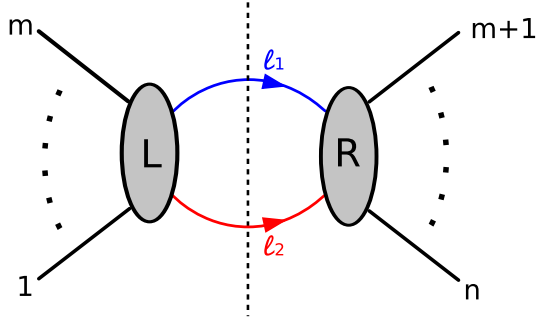}}
\caption{
\small\small\textsl{The cut of the one loop amplitude in the $t_1^{[m]}$ channel.}}\label{unitaritycut}
\end{figure}

We start by acting with the generator $\bar S_{1\to 1}$ on a finite cut of the one loop amplitude. To isolate the cut in a specific momentum channel $t_i^{[m]}=(k_i+\dots+k_{i+m-1})^2$, we consider the amplitude in the (unphysical) kinematical regime where $t_i^{[m]}>0$ and all other momentum invariants are negative. Without loss of generality, we assume that $i=1$ and the energy of $k_1+\dots+k_m$ is positive. In that kinematical regime, the discontinuity of the amplitude is computed by
\beq \label{cut}
\Delta_1^{[m]}\cA\equiv \cA(t_1^{[m]}+i0^+)-\cA(t_1^{[m]}-i0^+)=2i\,{\rm Im}\,\cA(t_1^{[m]}+i0^+)~.
\eeq
For the one loop amplitude, the result is given by (see figure \ref{unitaritycut})
\beqa \label{cutamp}
\Delta_1^{[m]}\cA_n^{(1)}= (2\pi)^2 \int d^4l_1d^4l_2\delta^{(+)}(l_1^2)\delta^{(+)}(l_2^2)\int d^4\eta_{l_1}d^4\eta_{l_2} \mathcal{A}_L \mathcal{A}_R
\eeqa
where
\beqa
\cA_L=\cA_{m+2}^{(0)}(l_1,l_2,\dots,m)~,\qquad \cA_R=\cA_{n-m+2}^{(0)}(-l_2,-l_1,\dots,n)~,\qquad\eta_{-l_{1,2}}=-\eta_{l_{1,2}}~.\nn
\eeqa
The finite cuts are the ones in multi-particle channels $2<m<n-2$ and in that section, we restrict our discussion to that range.

We now act on $\Delta_1^{[m]}\cA_n^{(1)}$ with the generator $\bar S^A_{\dot a}$ given in (\ref{treegen}). In the following, we will omit the indices from $\bar S$ since they can be trivially re-introduced. First, note that $\bar S$ fails to annihilate the discontinuity only due to the holomorphic anomaly. To see that we first  define
\beq\la{Baregen}
\bar S_L=\sum_{i=1}^ms_i\eta_i\bar\d_i~,\qquad\bar S_R=\sum_{i=m+1}^ns_i\eta_i\bar\d_i~,\qquad\bar S_{l_1,l_2}=\sum_{i=1}^2\eta_{l_i}\bar\d_{l_i}=-\sum_{i=1}^2\eta_{-l_i}\bar\d_{l_i}~.
\eeq
Then 
\begin{eqnarray}\label{total}
\bar S_{1\to 1}\Delta_1^{[m]}\cA_n^{(1)}=(2\pi)^2\int dLIPS(l_1,l_2)\int d^4\eta_{l_1}d^4\eta_{l_2}\[\(\bar S_L+\bar S_R+\bar S_{l_1,l_2}\)-\bar S_{l_1,l_2}\] \cA_L\cA_R~.
\end{eqnarray}
If we ignore the anomaly, the term in parentheses does not contribute because $(\bar S_L+\bar S_{l_1l_2})\cA_L=\bar S_R\, \cA_L=0$ with similar expressions for $\cA_R$.\footnote{Note that the internal momenta entering $\cA_R$ are $-l_1$ and $-l_2$. These have negative energy. However, since $\eta_{-l_{1,2}}=-\eta_{l_{1,2}}$, $\cA_R$ is annihilated by the sum $S_R+S_{l_1,l_2}$ and not by the difference $S_R-S_{l_1,l_2}$.} The last term, outside the parentheses, also vanishes since it is a total derivative
$$
\int d^4l\delta^{(+)}(l^2)\bar\d_l^{\dot a}f(l^{b\dot b})=\int_0^\infty dt\,t\int_{\widetilde\lambda=t\bar\lambda}\<\lambda_l,d\lambda_l\>[\widetilde\lambda_l,d\widetilde\lambda_l]{\d\over\d\widetilde\lambda_l}f((t\lambda_l^b)(\widetilde\lambda_l^{\dot b}))=0\,.
$$
We conclude that $\bar S\Delta_1^{[m]}\cA_n^{(1)}$ is non zero only due to the holomorphic anomaly. Moreover, for non collinear external momenta, it is supported on the region of integration where one of the internal momenta is collinear to one of the external momenta adjacent to the cut. For simplicity, we will only consider the case where the $n$ particle amplitude is MHV. In that case, both tree level sub amplitudes in (\ref{cutamp}) are MHV. Using the tree level MHV generating function (\ref{MHVgen}), acting with $\bar S$ and picking the contribution from the holomorphic anomaly, we find \cite{Bena:2004xu,Korchemsky:2009hm}
\begin{figure}[t]
\epsfxsize=15cm
\centerline{\epsfbox{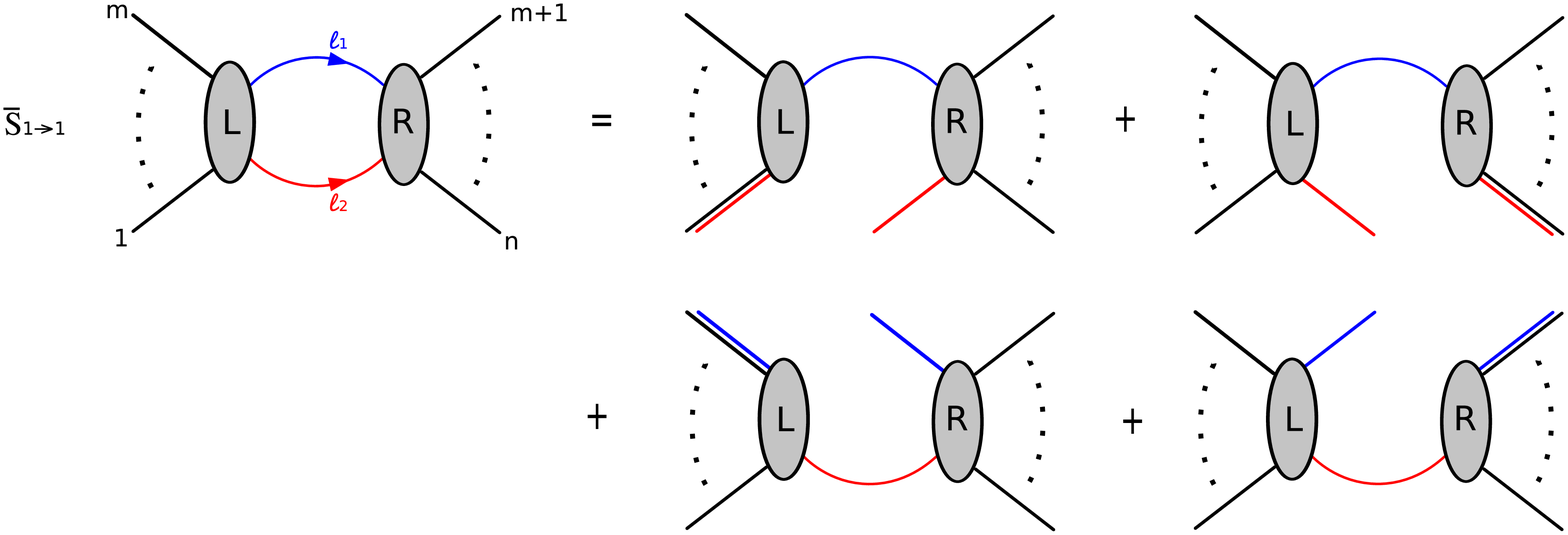}}
\caption{
\small\small\textsl{The action of the superconformal generator $\bar S_
{1\to 1}^{(0)}$ on a one loop unitarity cut. The holomorphic anomalies set an internal momenta to be collinear to an external one, giving rise to a $n+1$ tree level amplitude with two collinear particles. We deduce that the correction to that generator must be of the form $\bar S^{(0)}_{2\to 1}$.}}\label{figSonloop}
\end{figure}
\begin{eqnarray}\label{barScut}
\bar S_{1\to 1}\Delta_1^{[m]}\cA_n^{(1)}=
&-&i4\pi^3\[\eta_1P_L^2-\sum_{i=1}^m\eta_i\<i|\sP_L|1]\]\widetilde\lambda_1{\<m,m+1\>\theta(l_1^0)\theta(s_1x_1)\over\<m|\sP_L|1]\<m+1|\sP_L|1]}\cA_n^{(0)}\\
&+&i4\pi^3\[\eta_nP_L^2-\sum_{i=1}^m\eta_i\<i|\sP_L|n]\]\widetilde\lambda_n{\<m,m+1\>\theta(l_n^0)\theta(s_nx_n)\over\<m|\sP_L|n]\<m+1|\sP_L|n]}\cA_n^{(0)}\nn\\
&+&i4\pi^3\[\eta_mP_L^2-\sum_{i=1}^m\eta_i\<i|\sP_L|m]\]\widetilde\lambda_m{\<n1\>\theta(l_m^0)\theta(s_mx_m)\over\<n|\sP_L|m]\<1|\sP_L|m]}\cA_n^{(0)}\nn\\
&-&i4\pi^3\[\eta_{m+1}P_L^2-\sum_{i=1}^m\eta_i\<i|\sP_L|m+1]\]\widetilde\lambda_{m+1}{\<n1\>\theta(l_{m+1}^0)\theta(s_{m+1}x_{m+1})\over\<n|\sP_L|m+1]\<1|\sP_L|m+1]}\cA_n^{(0)}~,\nn
\end{eqnarray}
where $|i]$ stands for $\tilde \lambda_i=s_i \bar \lambda_i$ and
\beq\label{xiPl}
P_L=\sum_{i=1}^mk_i~,\qquad x_i={P_L^2\over 2k_i\cdot P_L}~,\qquad l_i=P_L-x_i k_i~.
\eeq
Notice that each line in (\ref{barScut}) has a clear origin, represented in figure \ref{figSonloop}. Namely, the first line comes from the holomorphic anomaly that sets $l_2$ and $\lambda_1$ to be collinear, i.e. it steams from the action of the superconformal generator on $1/\<1 l_2\>$.\footnote{The holomorphic anomaly that set $l_1$ collinear to $l_2$ do not contribute after preforming the Grassman integration over $\eta_{l_1}$ and $\eta_{l_2}$.} The other three lines, from top to bottom, come from the action on $1/\<l_2 n\>$, $1/\<m l_1\>$ and $1/\< l_1 m+1\>$. The relative signs originate from the sign difference between $1/\<\lambda_l \lambda_\alpha\>$ and $1/\<\lambda_\alpha \lambda_l\>$, where $l=l_1,l_2$ and $\alpha=1,m,m+1,n$.

The two step functions in each term restrict the energy of the two internal on-shell momenta to flow from the left to the right. In the kinematical regime we consider here, these step functions are automatically satisfied. That is because $l_1+l_2=P_L$ is a positive energy time-like momenta. We chose to write these step functions explicitly because latter they will be used for understanding the recipe for cutting $\bar S_{2\to 1}^{(0)}\cA^{(0)}$ in a general kinematical regime.

The calculation above is valid only for $2<m<n-2$. The cases $m=2,n-2$ deserves a more delicate treatment and will not be considered in this section. Next, we will show that the same expression (\ref{barScut}) is obtained from the $(n+1)$ tree level amplitude, by the action of
\beq\la{allgen}
\bar S_{2\to 1}=2\pi^2\sum_{s=\pm}s\int d^{4|4}\Lambda d^4\eta'd\alpha\bar\lambda\eta'\Tr\[J(\Lambda^s)\check J(\Lambda_1^s)\check J(\Lambda_2^s)\]~,
\eeq
where
\beq\label{Grassman}
\begin{array}{l}
\lambda_1=\lambda\sin\alpha\\
\lambda_2=\lambda\cos\alpha
\end{array}, \qquad \begin{array}{l}
\eta_1=\eta\sin\alpha-\eta'\cos\alpha \\
\eta_2=\eta\cos\alpha+\eta'\sin\alpha
\end{array}
\,.
\eeq
Notice that this is indeed the structure indicated by figure \ref{figSonloop}: we act on a $n+1$ tree level amplitude rendering two of its legs collinear.
Since in the previous section we restrict the one-loop $n$ amplitude to be MHV, we have to show that
\beq\label{toshow}
\Delta_1^{[m]}\[\bar S_{1\to 1}\cA_n^{(1)MHV}+\bar S_{2\to 1}^{+}\cA_{n+1}^{(0)NMHV}\]=0~,
\eeq
where the number of $\pm$'s stands for the change in the helicity charge (\ref{helicity}) with multiplicity of two\footnote{In (\ref{helicity}), $n$ is the number of particles and $4k$ is the number of $\eta$'s. Hence removing a leg reduce the helicity by 2 and integrating over $\eta'$ increase the helicity by 4.}. There are three other terms that in principle could have appeared: $\bar S_{1\to 1}^{(1)}\cA_n^{(0)MHV}$, $\bar S_{2\to 1}^-\cA_{n+1}^{(0)MHV}$ and $\bar S_{2\to 1}^{+++}\cA_{n+1}^{(0)N^2MHV}$, where $\bar S_{1\to 1}^{(1)}$ is a one loop correction of $\bar S_{1\to 1}$. The first one does not have a cut and the validity of (\ref{toshow}) means that $\bar S_{2\to 1}^-=\bar S_{2\to 1}^{+++}=0$.

\begin{figure}[t]
\epsfxsize=7cm
\centerline{\epsfbox{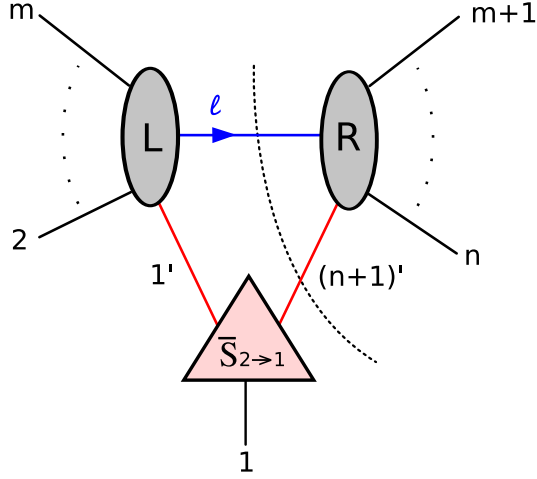}}
\caption{\small\small\textsl{The term in the CSW sum of $\cA^{(0)}_{n+1}$ that has a cut in $t_1^{[m]}=(k_1+k_2+\dots+k_m)^2$, when integrated over the collinearity portion of leg $1'$ and leg $(n+1)'$.}}\label{treecut}
\end{figure}

We would like to compare (\ref{barScut}) with the cut of $\bar S_{2\to 1}\cA_{n+1}^{(0)}$. The collinear $(n+1)$ amplitude is divergent. However, the cut of the generator is finite. That is because the divergent pieces of the collinear amplitude do not have a discontinuity and therefore drop out.
The action of $\bar S_{2\to 1}$ on $\cA_{n+1}^{(0)}$ produces a sum of terms in which one of the $n$-particles is replaced by two collinear adjacent particles. Only four of these terms have a discontinuity in the $t_1^{[m]}$ channel. These are the terms in which the particles adjacent to the cut  are -- $\{1,m,m+1,n\}$ -- see figure \ref{figSonloop}. Notice that these contributions are indeed finite. For simplicity, we isolate from the action of $\bar S_{2\to 1}$ on $\cA_{n+1}^{(0)}$ the term in which the two collinear momenta are collinear to $k_1$. The other terms are related to that by relabeling of the legs. We label the two collinear legs $1'$ and $(n+1)'$ to distinguish them from their sum $1=1'+(n+1)'$ which becomes leg $1$ of the $n$-particle amplitude. Using (\ref{allgen}) we find that the cut in the $t_1^{[m]}$ channel of that term is term is
\begin{eqnarray}
&& \nn \Delta_1^{[m]}\[\bar S_{2\to 1}\cA_{n+1}^{NMHV}\]_1=2\pi^2\Delta_1^{[m]}\int d^4\eta' \eta' \widetilde \lambda_1 \int_0^1 \frac{dx}{\sqrt{x(1-x)}} \cA_{n+1}^{NMHV}(1',\dots,(n+1)')\,,
\end{eqnarray}
where $x=\cos(\alpha)$. It is clear that we can move $\Delta_1^{[m]}$ freely into the integral and equally write
\begin{eqnarray}
&&\la{actionONcut} \[\bar S_{2\to 1}\Delta_1^{[m]}\cA_{n+1}^{(0)NMHV}\]_1=2\pi^2\int d^4\eta' \eta' \widetilde \lambda_1 \int_0^1 \frac{dx}{\sqrt{x(1-x)}} \Delta_1^{[m]}\cA_{n+1}^{(0)NMHV}(1',\dots,(n+1)')~\,.
\end{eqnarray}
Next, we express the tree level amplitude on the right hand side as a CSW sum\footnote{The same result can be obtained using BCFW \cite{BCFW} instead. Here, we chose to use CSW because it has a straightforward generalization to  loop level which we will use in the next section.} \cite{CSW}, i.e. as a sum over MHV vertices connected by off-shell propagators. As the $(n+1)$-amplitude at hand is NMHV, any term in the CSW sum consist of two MHV vertices connected by a single propagator. In the kinematical regime we are working, only one term has a discontinuity in the $t_1^{[m]}$ channel. That is the term (see figure \ref{treecut})
\beqa
\Delta_1^{[m]}\!\cA^{(0)}((1\!-\!x)1,2,\!\dots\!,n,x1) \la{almost}\!=\!\Delta_1^{[m]}\!\delta(P_n)\!\!\int \!\!d^4\eta_l {A^{(0)}((1\!-\!x)1,2,\!\dots\!,m,-l)A^{(0)}(l,m\!+\!1,\!\dots\!,n,x1)\over P_L^2-2xk_1\cdot P_L+i0}~,
\eeqa
where
$$
(\lambda_l)_a=(P_L-xk_1)_{a\dot a}\chi^{\dot a}
$$
and $\chi$ is an arbitrary fixed null vector. We have removed the subscripts $n+1$, $m+1$ and $n-m+2$ since they can be easily read off by counting the number of arguments of the corresponding amplitude. Using (\ref{cut}) and the relation
\beq
{i\over y+i0}-{i\over y-i0}=2\pi \delta(y)~. \la{delta}
\eeq
we simplify $\Delta_1^{[m]}\!\cA^{(0)}((1\!-\!x)1,2,\!\dots\!,n,x1)$ in (\ref{almost}) to
\beqa
2\pi i\delta(P_n){\rm sign}(P_L\cdot k_1)\delta(x-x_1)\int d^4\eta_l A^{(0)}((1-x)1,2,\dots,m,-l_1){x\over P_L^2}A^{(0)}(l_1,m+1,\dots,n,x1)~,\label{CSWterm}
\eeqa
where $x_1$ and $l_1$ are given in (\ref{xiPl})\footnote{Note that the dependence on $\chi$ has dropped out.}. In the kinematical regime we are working in, $P_L^2>0$. For $x_1\in[0,1]$ it means that ${\rm sign}(P_L\cdot k_1)=1$. Next, we plug (\ref{CSWterm}) back into (\ref{actionONcut})
\beqa
&&\[\bar S_{2\to 1}\Delta_1^{[m]}\!\cA_{n+1}^{(0)NMHV}\]_1\label{cutbarStree}\\&=&\!\!\!\!i4\pi^3{\widetilde \lambda_1\over P_L^2}\int_0^1 dx\delta(x-x_1)\sqrt{\!{x\over 1-x}}\!\int\!\! d^4\eta_l d^4\eta' \eta'  A^{(0)}((1\!-\!x)1,2,\!\dots\!,m,-l)A^{(0)}(l,m\!+\!1,\!\dots\!,n,x1)\nn\\&=&\!\!\!\!i4\pi^3\cA_n^{(0)}\[\eta_1P_L^2-\sum_{i=1}^m\eta_i\<i|\sP_L|1]\]\widetilde\lambda_1{\<m,m+1\>\over\<m|\sP_L|1]\<m+1|\sP_L|1]}~,\nn
\eeqa
where in the last step we preformed the Grassmanian integration using (\ref{Grassman}). Note that in the kinematical regime we are working $x_1=t_1^{[m]} /( t_1^{[m]}-t_2^{[m-1]})\in [0,1]$ is always inside the region of integration. The final result in \rf{cutbarStree} is exactly minus the first line of (\ref{barScut})! A summation over the three other term corresponding to particles $m$, $m+1$ and $n$ reproduces (\ref{barScut}) and confirms (\ref{toshow}).

For (\ref{toshow}) to hold in a general kinematical regime, we must reproduce the step functions in (\ref{barScut}). Physically, these step function restrict the energy flow in the cut lines of figure \ref{unitaritycut} to flow from the left to the right. The $\theta(l_1^0)$ is reproduced by cutting the tree level propagator between the two MHV vertices
$$
{1\over L^2}\quad\rightarrow\quad \delta^{(+)}(L^2)~.
$$
The second step function $\theta(s_1x_1)$, has to be associated with the procedure of cutting a leg connecting $\cJ_{2\to 1}$ to the amplitude. It restrict the energy component of the corresponding collinear particle to be positive (see Fig \ref{treecut}). We suggest it to be the general procedure for taking unitarity cuts of $\cJ_{n \to m} \cA$.

\section{Formal Superconformal Invariance of One Loop MHV Amplitudes}\la{sec:formal}

In the previous section we demonstrated the superconformal invariance of unitarity cuts. In this section we will show that a formal invariance continues to hold for the full one loop MHV amplitude. The invariance will only be formal because some of the integrals we will consider are divergent. That is however not the first time where non-trivial information is obtained from formal manipulations of divergent integrals. For example, in \cite{Bern:1994zx} Bern, Dixon, Dunbar and Kosower computed the one loop MHV amplitudes of $\cN=4$ SYM by formal manipulations of its unitarity cuts. What allowed them to do so was the independent knowledge that these amplitude are given by a sum of box integrals. In our case the logic is different. That is, we will use these formal manipulations to define the superconformal generators. Then, in section \ref{sec:reg}, we will show that up to a conformal anomaly, the structure survives regularization.    

We start by repeating the computation of $\bar S_{2\to 1}\cA_{n+1}^{(0)}$ above but without taking its unitarity cut. That is, we formally remove the cut from (\ref{actionONcut})
\beq\label{Nocut}
\[\bar S_{2\to 1}\cA_{n+1}^{(0){\rm NMHV}}\]_1=2\pi^2\int d^4\eta' \eta' \widetilde \lambda_1 \int_0^1 \frac{dx}{\sqrt{x(1-x)}} \cA_{n+1}^{(0){\rm NMHV}}(1',\dots,(n+1)')
\eeq
and represent the tree amplitude as a CSW sum \cite{CSW}
\beq\label{CSWsum}
\cA_{n+1}^{(0){\rm NMHV}}(1,\dots,n+1)=-i\sum_{i=1}^{n+1}\sum_{m=2}^{n-1}\int d^4\eta_l\int {d^4L\over L^2}\widetilde\cA_L^{(0){\rm MHV}}\widetilde\cA_R^{(0){\rm MHV}}~,
\eeq
where
\begin{eqnarray}
\widetilde\cA_L^{(0){\rm MHV}}&=&\delta^4(P_L+L) A_{m+1}^{(0){\rm MHV}}(l,i,\dots,i+m-1) \,,  \la{ALRtree}\\
\widetilde\cA_R^{(0){\rm MHV}}&=&\delta^4(P_R-L) A_{n-m+1}^{(0){\rm MHV}}(-l,i+m,\dots,i-1)~,\nn
\end{eqnarray}
and
\beq\label{PLPRl}
P_L=\sum_{r=i}^{i+m-1}k_r \,,\qquad P_R=\sum_{r=i+m}^{i-1}k_r \, ,\qquad l=L-y\chi~,\qquad y={P_L^2\over 2P_L\cdot\chi}~,
\eeq
where $\chi$ is an arbitrary null vector. The only difference between the $\widetilde\cA^{(0){\rm MHV}}$ and tree level MHV amplitudes $\cA^{(0){\rm MHV}}$ is in the momentum conservation delta function where the off-shell momentum $L$ enters and not the on-shell momenta $l$.\footnote{Not only the off-shell amplitudes $\widetilde\cA_n^{(0){\rm MHV}}$ are still annihilated by $\bar S_{1\to 1}$ for generic external momenta but also the correction $\bar S_{1\to 2}$ required to account for collinear external momenta comes from the same holomorphic anomaly and is therefore trivially generalized to act on these amplitudes. The generators $\bar S_{1\to 1}$ and $\bar S_{1\to 2}$ generalized to act on these amplitudes will be given latter in (\ref{offS11}) and (\ref{offS12}).}
The superconformal generator in (\ref{Nocut}) sets two of the momenta in (\ref{CSWsum}) to be collinear; these two momenta can belong to different sub-amplitudes (one in $\mathcal{A}_L$, the other in $\mathcal{A}_R$) or they can be on the same side (both in $\mathcal{A}_L$ or both in $\mathcal{A}_R$). Terms where the two collinear momenta are on the same side vanish when plugged into (\ref{Nocut}). That is because these terms are proportional to $\<1'(n+1)'\>^{-1}$ whereas the Grassman integral over $\eta'$ produces a factor of $\<1'(n+1)'\>^3$ (resulting in a total factor of $\<1'(n+1)'\>^2$).

Of course (\ref{CSWsum}) can be simplified using
$$
\int{d^4L\over L^2}\delta^4(P_L+L)\delta^4(P_R-L)={\delta^4(P_L+P_R)\over P_L^2} \,.
$$
We will not do so here but instead express it as \cite{Brandhuber:2004yw}
\beq\label{onshelldecomposition}
\int{d^4L\over L^2}\delta^4(P_L+L)=\int \frac{dy}{y} \int d^4 l\, \delta(l^2)\,\delta^{(4)}(P_L+y \chi+l){\rm sign}(\chi\cdot l) \,.
\eeq
Preforming the Grassman integrations over $\eta_l$ and $\eta'$, we get
\beqa\nn
\[\bar S_{2\to 1}\cA_{n+1}^{(0){\rm NMHV}}\]_1&=&i2\pi^2\widetilde\lambda_1\cA_n^{(0)}\sum_{m=2}^{n-1} \int_0^1 dx x  \int \frac{dy}{y} \int d^4l\, \delta(l^2) \delta^{(4)}(P_{L,y}-xk_1 - l){\rm sign}(\chi\cdot l)\\&\times& {1\over P_{L,y}^2}\[\eta_1P_{L,y}^2-\sum_{j=1}^m\eta_j\<j | \sP_{L,y} |1 ]\]\frac{\<l1\>^2\<m,m+1\>}{\<m l\> \< l,m+1\>}~.\nn
\eeqa
where\footnote{When acting with the superconformal generator, the momenta $k_1 \in P_L$ becomes $k_{1'}=(1-x)k_1$ hence justifying the extra term $-x k_1$ inside the delta function.}
\beq
P_{L,y}=k_1+\dots + k_m-y \chi~.
\eeq
We can now integrate over $y$ and $l$ to obtain
\beq\label{barSontree}
\hspace{-2mm}
\boxed{
\[\bar S_{2\to 1}\cA_{n+1}^{(0){\rm NMHV}}\]_1=i2\pi^2\widetilde\lambda_1\cA_n^{(0)}\sum_{m=2}^{n-1}\int_0^1{dx\over x}{P_{L,y}^2\over P_{L,x}^2}
\[\eta_1P_{L,y}^2-\sum_{j=1}^m\eta_j\<j | \sP_{L,y} |1]\]\frac{\<m,m+1\>}{\<m| \sP_{L,y} |1] \<m+1| \sP_{L,y} | 1]}}~,
\eeq
where\footnote{Note that $\int{dx\over x}{P_{L,y}^2\over P_{L,x}^2}=\int{dy\over y}$.}
$$
y={P_{L,x}^2\over 2P_{L,x}\cdot \chi}~,\qquad P_{L,x}=k_1+\dots + k_m-xk_1~.
$$

Before we move on and consider the action of the superconformal generators on the one loop amplitude, a few comments are in order:
\begin{enumerate}
\item[$\bullet$] Taking the cut of (\ref{barSontree}) in the $t_1^{[m]}$ channel localizes the $y$-integral at $y=0$, yielding (\ref{cutbarStree}).
\item[$\bullet$] Any term in the sum depends on the arbitrary chosen null vector $\chi$. The sum is however $\chi$ independent.
\item[$\bullet$] For compactness of the expressions above, we have dropped the explicit $i\epsilon$ prescription of the Feynman propagator. It is trivial to add it back as will be done below.
\item[$\bullet$] For $m=1,2,n-2,n-1$ the integrals in (\ref{barSontree}) are divergent. In section \ref{sec:reg}   we will deal with their regularization.
\end{enumerate}

Next, we would like to compare (\ref{barSontree}) with the action of $\bar S_{1\to 1}$ on the $n$-particle MHV amplitude $\cA_n^{(1){\rm MHV}}$. In \cite{Brandhuber:2004yw}, a generalization of the CSW formula to one-loop MHV amplitudes was given as
\beq
\cA_n^{(1){\rm MHV}}=-2\pi i\sum_{i=1}^n\sum_{m=1}^{n-1}\int {dy\over y+i0}\int d^4l_1d^4l_2\delta^{(+)}(l_1^2)\delta^{(+)}(l_2^2)\int d^4\eta_{l_1}d^4\eta_{l_2}\widetilde\cA_L^{(0){\rm MHV}}\widetilde\cA_R^{(0){\rm MHV}}~, \la{oneloopcut}
\eeq
where
\begin{eqnarray}\label{PLPR}
\widetilde\cA_L^{(0){\rm MHV}}&=&\delta^4(P_L-l_1-l_2-y\chi) A_{m+1}^{(0){\rm MHV}}(-l_1,-l_2,i,\dots,i+m-1) \,,  \\
\widetilde\cA_R^{(0){\rm MHV}}&=&\delta^4(P_R+l_1+l_2+y\chi) A_{n-m+1}^{(0){\rm MHV}}(l_2,l_1,i+m,\dots,i-1)\,,\nn
\end{eqnarray}
the left and right momenta $P_L$, $P_R$ are given in (\ref{PLPRl}) and we have chosen $\chi$ to have positive energy ($\chi^0>0$).\footnote{The step functions $\theta(l_1^0)\theta(l_2^0)$ imply that $P_L-y \chi$ is the sum of two positive energy null momenta and must therefore be a (positive energy) time-like vector. Thus, the integrand in each of the summands in (\ref{oneloopcut}) is nonzero for $y\ge -\frac{P_L^2}{2\chi\cdot P_L}$ (for $m=1$ this yields $y\ge 0$).}

For any fixed value of $y$,  the dLIPS integral in (\ref{oneloopcut}) computes the discontinuity of a one loop amplitude in the $P_{L,y}$ channel (where the tree level amplitude has been factored out). It depends on $y\chi$ only through the momentum conservation delta functions in (\ref{PLPR}). We can therefore apply the result of section \ref{sec:fincuts} directly to the loop amplitude. As before, $\bar S_{1\to 1}$ fails to annihilate the one loop amplitude only due to the holomorphic anomaly and we isolate the term in which an internal on-shell momenta is collinear to $k_1$
\beq\label{barSonloop}
\hspace{-2mm}
\boxed{
\[\bar S_{1\to 1}\cA_n^{(1){\rm MHV}}\]_1=-i2\pi^2\widetilde\lambda_1\cA_n^{(0)}\sum_{m=2}^{n-1}\int_0^1{dx\over x}{P_{L,y}^2\over P_{L,x}^2}\[\eta_1P_{L,y}^2-\sum_{j=1}^m\eta_j\<j | \sP_{L,y} |1]\]\frac{\<m,m+1\> }{\<m| \sP_{L,y} |1] \<m+1| \sP_{L,y} | 1]}}~,
\eeq
where
$$
x={P_{L,y}^2\over 2P_{L,y}\cdot k_1}~.
$$
Comparing with (\ref{barSontree}) we see that, at the level of formal integrals, we obtain a match between the two expression, i.e.,
\beq\label{FormalSummary}
\bar S_{1\to 1}\cA_n^{(1){\rm NMHV}}+\bar S_{2\to 1}\cA_{n+1}^{(0){\rm NMHV}}=0~.
\eeq
In obtaining (\ref{barSonloop}) there were two point that deserve explanation.
\begin{enumerate}
\item[$\bullet$] It is quite nontrivial that we obtain precisely the same region of integration $0<x<1$ in (\ref{barSonloop}) and in (\ref{barSontree}). Each $m$ summand in (\ref{barSonloop}) originates from four terms in the action of $\bar S_{1\to 1}$ on (\ref{oneloopcut}). These are the terms in which $(i,m)$ in (\ref{oneloopcut}) are equal to $\{(1,m),(m+1,n-m),(2,m-1),(m+1,n-m+1)\}$, where the first two are drawn in figure \ref{BarS21}.a and the last two in figure \ref{BarS21}.b. Each of these four terms is the same as (\ref{barSonloop}) with the integrand multiplied by the following step functions (see (\ref{barScut}))
\begin{figure}[t]
\epsfxsize=16cm
\centerline{\epsfbox{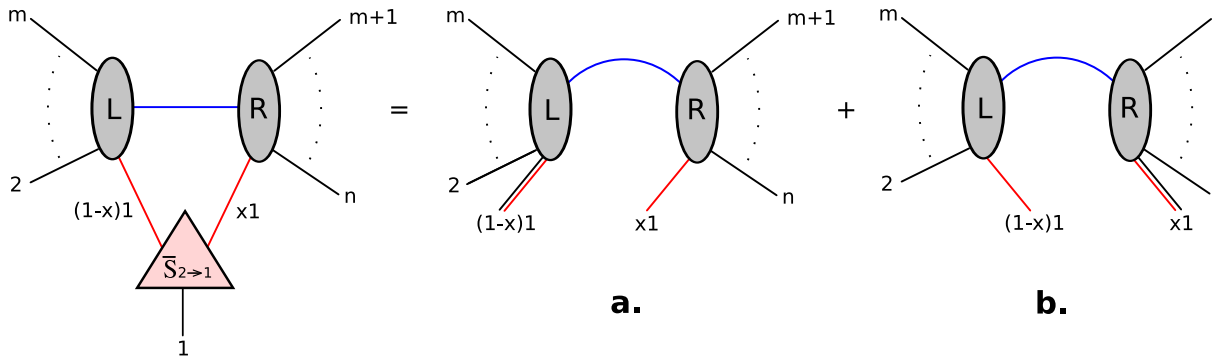}}
\caption{\small\small\textsl{{\bf a} and {\bf b} are terms that appear in the action of $\bar S_{1\to 1}^{(0)}$ on the one loop MHV $n$-particle amplitude $\cA_n^{(1)}$, see figure \ref{figSonloop}. They correspond to two different terms in the CSW sum of the loop amplitude where an internal on-shell momenta become collinear to $k_1$. Their sum is equal to a term in $\bar S_{2\to 1}\cA_{n+1}^{(0)}$ where the two collinear particles are collinear to $k_1$. That term is drawn on the left.}}\label{BarS21}
\end{figure}
\beqa\la{Regions}
(1,m):&\qquad&+s_1\theta(+(P_{L,y}^0-xk_1^0))\theta(+xk_1^0)\nn\\
(m+1,n-m):&\qquad&-s_1\theta(-(P_{L,y}^0-xk_1^0))\theta(-xk_1^0)\\
(2,m-1):&\qquad&-s_1\theta(+(P_{L,y}^0-xk_1^0))\theta(-(1-x)k_1^0)\nn\\
(m+1,n-m+1):&\qquad&+s_1\theta(-(P_{L,y}^0-xk_1^0))\theta(+(1-x)k_1^0)~.\nn
\eeqa
When summing these four terms the dependence on $P_L$ and $s_1$ drops out leaving just
\beq\la{Cancelation}
\theta(x)\theta(1-x)~.
\eeq
as in (\ref{barSonloop}). The cancelation of the $P_L$ dependence in the region of the $x$ integration is essential for the locality of the correction $\bar S_{2\to1}$.
Notice that for (\ref{FormalSummary}) to hold, it was crucial that in the definition of $\bar S_{2\to1}$  the portion of collinearity $x$ was integrated only between 0 and 1. Physically this means that the two collinear particles have the same energy sign.

\item[$\bullet$] Note that the sum in (\ref{barSonloop}) starts at $m=2$. On the other hand, the CSW one loop sum (\ref{CSWsum}) contains a term where $(m,i)=(1,1)$. That is the term where one of the MHV vertices is a three vertex connecting particle $1$ to the two internal propagators. It already contributes to (\ref{barSonloop}) from the regions of integration where $l_1$ become collinear to $k_2$ or $l_2$ becomes collinear to $k_n$. One may expect that it will also contribute to (\ref{barSonloop}) from the region of integration where an internal momentum becomes collinear to $k_1$. In that term, the momentum conservation delta function of the MHV three vertex is $\delta^4(k_1-y\chi-l_1-l_2)$. Suppose $l_1=tk_1$. The only way to have momentum conservation is if $y=0$ or $t=1$.\footnote{That is, generically the holomorphic anomaly is supported outside the region of integration.} The point $t=1$ is however a point of measure zero in the dLIPS integration. The corresponding holomorphic anomaly is therefore supported at the point where $y=0$. That is the point where the original $y$-integrand is divergent. A formal manipulation of that contribution is therefore invalid. In other words, to make any sense out of that contribution, we must first introduce a regulator. After regularization, that contribution vanish. As the details are technical and involve a regularization not yet introduced, we present them in Appendix A.

\end{enumerate}

\section{Generalizations to All Loops and Helicities} \la{all}
In \cite{Brandhuber:2004yw,Brandhuber:2005kd}  a generalization of the CSW formula to one-loop MHV amplitudes was given as\footnote{The relation between (\ref{oneloopCSW}) and (\ref{oneloopcut}) will be reviewed in detail in the next section.}
\beq \label{oneloopCSW}
\cA_n^{(1){\rm MHV}}=-\sum_{i=1}^n\sum_{m=1}^{n-1}\int{d^4L_1\over L_1^2+i0}\int{d^4L_2\over L_2^2+i0}\int d^4\eta_{l_1}d^4\eta_{l_2}\widetilde\cA_{L}\widetilde\cA_{R}~,
\eeq
where
\begin{eqnarray}\label{PLPRCSW}
\widetilde\cA_L&=&\delta^4(P_L+L_1+L_2) A_{m+1}^{(0){\rm MHV}}(l_2,l_1,i,\dots,i+m-1)  \\
\widetilde\cA_R&=&\delta^4(P_R-L_1-L_2) A_{n-m+1}^{(0){\rm MHV}}(-l_1,-l_2,i+m,\dots,i-1)\nn
\end{eqnarray}
and
\beq\la{offshelltonull}
L_i=l_i+y_i\chi~.
\eeq
Here $\chi$ is an arbitrary chosen null vector.  For every fixed values of $i$ and for every $m$ in the sum we can express the $L_1$ integral as
\beqa\la{Ltol}
\int{d^4 L_1\over L_1^2+i0}\delta^4(P_{L_m}+ L_1) =\int d^4l_1\delta(l_1^2){\rm sign}(l_1^0)\int{dy\over y+i0\ {\rm sign}(l_1^0)}\delta^4(P_{L_m}+l_1+y\chi)~,\nn
\eeqa
where
$$
P_{L,m}=L_2+\sum_{j=i}^{i+m-1}k_j
$$
and the energy sign of $\chi$ was chosen to be positive. Now, for every fixed value of $i$, $y$ and $l_1$, the sum over $m$ reproduce the CSW formula for an $(n+2)$ tree level (off-shell continued) NMHV amplitude $\widetilde\cA_{n+2}^{(0){\rm NMHV}}$ with two adjacent legs been $l_1$ and $-l_1$ with momentum insertions $l_1+y\chi$ and $-l_1-y\chi$ correspondently\footnote{To be more precise, the CSW tree level sum also includes terms where the legs $l_1$ and $-l_1$ are attached to the same MHV vertex. These terms are proportional to the tree level splitting function that diverge as $1/\<l_1,-l_1\>$. However, the Grassmanian integration over $\eta_{l_1}$ produces a factor of $(|l_1\>-|-l_1\>)^4$, killing these terms \cite{Brandhuber:2005kd}. Note that even if we multiply first by $\eta_{l_1}^A$, these terms will not contribute.} (when expressing $\widetilde\cA_{n+2}^{(0){\rm NMHV}}$  as a CSW sum, the null momenta used to go off-shell must be $\chi$ and cannot be chosen independently). We have then
\beqa\nn
\cA_n^{(1){\rm MHV}}\!=\!-i\sum_{i=1}^n\int\! d^4l\delta(l^2){\rm sign}(l^0)\!\int\!{dy\over y+i0\ {\rm sign}(l^0)}\!\int \!d^4\eta_l\widetilde\cA_{n+2}^{(0){\rm NMHV}}\![(l,y\chi),(-l,-y\chi),i,\dots,i+n-1]\nn
\eeqa
where $\widetilde\cA^{(0)}$ is the off-shell continuation of tree-level amplitudes by means of momentum conservation at MHV vertices.  That is, the external momenta entering $\widetilde\cA$ can be off-shell and are treated in the same way as internal off-shell momenta via the CSW prescription \cite{CSW}.

We can replace the ${\rm sign}(l_0)$ in this expression by explicitly summing over positive and negative energy momenta $l$ thus obtaining
\beqa\la{looptree}
\cA_n^{(1){\rm MHV}}=-i\sum_{i=1}^n\sum_{s=\pm} \int{dy\over y+i0}\int {d^{4|4}\Lambda\over2\pi}\widetilde\cA_{n+2}^{(0){\rm NMHV}}[(sl,sy\chi),(-sl,-sy\chi),i,\dots,i+n-1]~,
\eeqa
where $d^{4|4}\Lambda=d^4l\delta^{(+)}(l^2)  d^4\eta_ld\varphi$.

We will now use the above observation to conjecture a generalization of CSW \cite{CSW} to any loop order and any helicity configuration (not necessarily MHV). For that aim, we first introduce a couple of definitions
\begin{itemize}
\item We define the generating function of all generalized MHV vertices
$$
\widetilde\cA^{(0)\rm MHV}_c[J]=\sum_{n=3}^\infty{g^{n-2}\over n}\int\prod_{i=1}^n d^{4|4}\Lambda_i dy_i\Tr[J(\widetilde\Lambda_{1,y_1})\dots J(\widetilde\Lambda_{n,y_n})]\widetilde\cA^{(0){\rm MHV}}_n(\Lambda_{1,y_1},\dots,\Lambda_{n,y_n})~,
$$
where $\widetilde\Lambda^\pm_y=(\Lambda^\pm,y)$ and
$$
\widetilde\cA^{(0){\rm MHV}}_n(\Lambda_{1,y_1},\dots,\Lambda_{n,y_n})=\delta^4\(\sum_{i=1}^n(k_i+y_i\chi)\)A_n^{(0)\rm MHV}(\Lambda_1,\dots,\Lambda_n)~.
$$
\item Next, we define the ``propagator inserting operator"
\beqa\la{loopraising}
\cL=i\int{dy \over y+i0}\int {d^{4|4}\Lambda\over2\pi}\Tr[\check J(\widetilde\Lambda_{y})\check J(-\widetilde\Lambda_y)]~,\nn
\eeqa
where $\check J(\widetilde\Lambda_y)={\delta\over\delta J(\widetilde\Lambda_y)}$ and $(-\widetilde\Lambda_y)=(\lambda,-\bar\lambda,-\eta,-y)$.
\item Finally, we express the full $\cN=4$ S-matrix $\bS_{matrix}$ (generating all connected, disconnected, planar and non-planar amplitudes) as
\beq
\bS_{matrix}[J]=e^{F[J]}\bS[J]~, \la{Smatrix}
\eeq
where \cite{Kim:1996nd}
\beq
F[J]=\int d^{4|4}\Lambda\Tr[J(\Lambda^+) \hat J(\Lambda^-)] \la{eqF}
\eeq
is introduced to take into account from sub-processes where some of the particles fly by unscattered and $\bS$ is the interacting part of the S-matrix. It is equal to the exponent of all connected amplitudes with three or more external particles.\footnote{Note that $\bS$ is not the transfer matrix (the latter only excludes the process where \textit{all} particles fly by unscattered).}  
\end{itemize}

Using these definitions, the conjectured CSW generalization reads
\beq\la{CSWloop}
\hspace{-2mm}
\boxed{
\bS[J]=e^{\cL}~e^{\widetilde\cA^{(0)\rm MHV}_c[J]}}~.
\eeq
Note that at tree level and for one loop MHV amplitudes this is \textit{not} a conjecture, see respectively \cite{BCFW,Risager:2005vk} and \cite{Brandhuber:2004yw}.\footnote{The CSW construction was argued to hold for any helicity configuration in \cite{Brandhuber:2005kd}. Furthermore,the existence of an MHV Lagrangian which is obtained from the usual one by a field redefinition after light-cone gauge fixing \cite{axial} provides additional strong evidence towards the exactness of this expansion,see \cite{Ettle:2007qc} where the (non-local) field redefinitions were argued to be mild enough not to raise any issues at both tree level and one loop.}
We will now study the symmetry transformations of this object; the transformation properties of the full S-matrix $\bS_{matrix}$ will then be read off from these.
\begin{figure}[t]
\epsfxsize=16cm
\centerline{\epsfbox{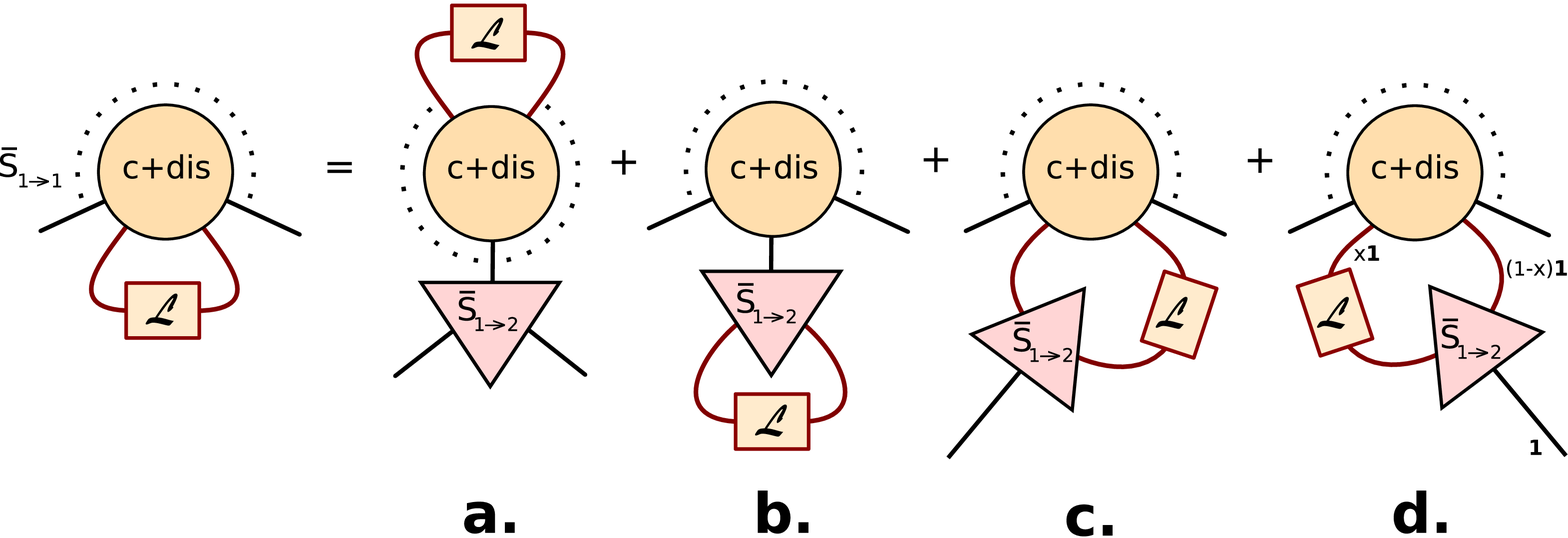}}
\caption{\small\small\textsl{
Result of the action of $\bar S_{1\to 1}$ on $\bS^{(1)}$. The operator $\bar S_{1\to 1}$ goes through $\cL$ thus acting on the MHV tree level generating function $\bS^{(0)}$. From \cite{Bargheer:2009qu} this gives rise to the action of $\bar S_{1\to 2}$ on  $\bS^{(0)}$. The two legs created by $\bar S_{1\to 2}$ can (a) be unrelated to the legs on which $\mathcal{L}$ acts, thus yielding terms which vanish for generic external momenta, (b) be acted upon by $\mathcal{L}$, giving a vanishing contribution due to the Grassmanian integration or (c,d) one of them can become an external leg while the other is acted upon by $\mathcal{L}$. The latter two contributions are identified with $\bar S_{2\to 1}$, (see figure \ref{Blue}).
}}\label{LoopOperator}
\end{figure}
We start by writing a recursive relation for the number of propagators between CSW MHV vertices. To do so,  we first introduce a parameter $x$ counting the number of such propagators
$$
\bS[x,J]=e^{x\cL}~e^{\widetilde\cA^{(0)\rm MHV}_c[J]}=\sum_{m=0}^\infty x^m\bS^{(m)}[J]~.
$$
To obtain the S-matrix, we set $x=1$
$$
\bS[J]=\bS[1,J]~.
$$
We can now write a recursive relation these coefficients\footnote{Here and everywhere the tilde stands for generalized off-shell amplitudes in the CSW sense.}
\beq\la{recursive}
\bS^{(m)}[J]={\cL\over m} \widetilde \bS^{(m-1)}[J]~.
\eeq
Let us now explain how we can recover and generalize our previous results assuming (\ref{CSWloop}). First, we note that for $\bar S$, the results of \cite{Bargheer:2009qu} applies as well for generalized tree level MHV amplitudes (these are the CSW vertices)\footnote{Note that MHV amplitudes don't have multi-particle poles.}
\beq\la{firststep}
\(\bar S_{1\to1}+g\bar S_{1\to2}\)\bS^{(0)}[J]=0~,
\eeq
where for generalized off-shell legs
\beqa
\bar{{S}}_{{1\to 1}}&=&-\sum_{s=\pm}s\int d^{4|4}\Lambda\,dy\, \eta  \,  \Tr\[\bar\partial J(\widetilde\Lambda^s_y) \check{J}(\widetilde\Lambda^s_y)\]  ~, \la{offS11} \\
\bar S_{1\to2}&=&2\pi^2{\sum_{s,s_1,s_2=\pm}}\!\!\!\!\!'~s\int dy_1dy_2\int d^{4|4}\Lambda d^4\eta'd\alpha\bar\lambda\eta'\Tr\[\check J(\widetilde\Lambda^s_{y_1+y_2})\hat J(\widetilde\Lambda_{1,y_1}^{s_1}) J(\widetilde\Lambda_{2,y_2}^{s_2})\]~. \la{offS12}
\eeqa
Now we recursively act with the bare generator $\bar S_{1\to1}$ on $\bS^{(1)}[J]$ using (\ref{recursive})
\beqa
\bar S_{1\to1}\bS^{(1)}[J]&=&\bar S_{1\to1}\cL\widetilde \bS^{(0)}[J]=\cL\bar S_{1\to1}\widetilde \bS^{(0)}[J]=-g\cL\bar S_{1\to2}\widetilde \bS^{(0)}[J]\nn\\&=&-g\bar S_{1\to2}\widetilde \bS^{(1)}[J]-g[\cL,\bar S_{1\to2}]\widetilde \bS^{(0)}[J]~,\nn
\eeqa
where when commuting $\cL$ through $\bar S_{1\to1}$, we used the fact that $l$ is not an external leg and that
\beq\la{Sbarl}
\bar S_l=\eta_l{\d\over\d\widetilde\lambda_l}=\eta_{-l}{\d\over\d\widetilde\lambda_{-l}}=\bar S_{-l}
\eeq
is a total derivative.
\begin{figure}[t]
\epsfxsize=10cm
\centerline{\epsfbox{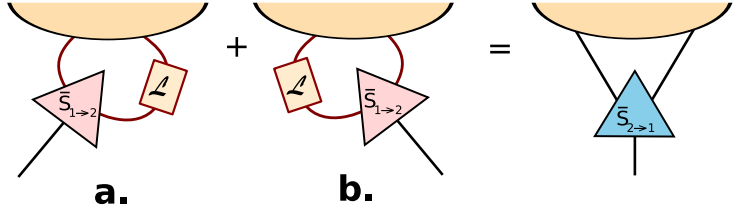}}
\caption{\small\small\textsl{The combined action of the propagator inserting operator $\mathcal{L}$ and the leg splitting operator $\bar S_{1\to 2}$ gives rise to the leg joining operator $\bar S_{2\to 1}$.}}\label{Blue}
\end{figure}
Now, (see figure \ref{Blue})
\beqa\la{UsingBiesert}
[\cL,\bar S_{1\to2}]&=&i\pi\int{dy\over y+i0}{\sum_{s,s_1,s_2=\pm}}\!\!\!\!\!'~s\int d^{4|4}\Lambda d^4\eta'd\alpha\bar\lambda\eta'\Tr\[\check J(\widetilde\Lambda^s_{s_1y})\check J(-\widetilde\Lambda^{s_1}_{1,s_1y}) J(\Lambda_2^{s_2})\]\\&+&i\pi\int{dy\over y+i0}{\sum_{s,s_1,s_2=\pm}}\!\!\!\!\!'~s\int d^{4|4}\Lambda d^4\eta'd\alpha\bar\lambda\eta'\Tr\[\check J(-\widetilde\Lambda^{s_2}_{2,s_2y})\check J(\widetilde\Lambda^s_{s_2y}) J(\Lambda_1^{s_1})\]~,\nn
\eeqa
The two terms in (\ref{UsingBiesert}) comes from contracting the right $(1)$ and the left $(2)$ legs of $\bar S_{1\to 2}$ with a neighboring leg using $\mathcal{L}$. These are drawn in figures \ref{Blue}.a and \ref{Blue}.b respectively. By the following change of variables in the last line
$$
(s,s_1,s_2,y)\quad\rightarrow\quad(-s_1,s_2,-s,ss_1y)~,
$$
the sum of the two terms can be rewritten as
\beqa\la{intygen}
[\cL,\bar S_{1\to2}]
=\pi{\sum_{s,s_1,s_2=\pm}}\!\!\!\!\!'~s\int dy\[{i\over y+i0}-{i\over y+iss_10}\]\int d^{4|4}\Lambda d^4\eta'd\alpha\bar\lambda\eta'\Tr\[\check J(\widetilde\Lambda^s_{s_1y})\check J(-\widetilde\Lambda^{s_1}_{1,s_1y}) J(\Lambda^{s_2}_2)\]~.\nn
\eeqa
For $s=s_1$ the two terms cancel. That is the same cancelation obtained in (\ref{Cancelation}). For $s=-s_1$ (and therefore $s=s_2$), the two terms gives a delta function (\ref{delta}).
We conclude that
\beq\la{corrAssume}
\(\bar S_{1\to1}+g\bar S_{1\to 2}\)\bS^{(1)}[J]+g\bar S_{2\to1}\bS^{(0)}[J]=0~,
\eeq
where
\beq
\hspace{-2mm}
\boxed{
\bar S_{2\to1}=2\pi^2\sum_{s=\pm}s\int d^{4|4}\Lambda d^4\eta'd\alpha\bar\lambda\eta'\Tr\[J(\Lambda^s)\check J(\Lambda_1^s)\check J(\Lambda_2^s)\]}~,
\eeq
is independent of $\chi$ and
\beq\label{Grassman2}
\begin{array}{l}
\lambda_1=\lambda\sin\alpha\\
\lambda_2=\lambda\cos\alpha
\end{array}, \qquad \begin{array}{l}
\eta_1=\eta\sin\alpha-\eta'\cos\alpha \\
\eta_2=\eta\cos\alpha+\eta'\sin\alpha
\end{array}~.
\eeq
This is precisely the form of the generator \rf{allgen} derived in the previous sections from the direct action on one loop MHV amplitudes.
Next, we use (\ref{recursive}) and (\ref{corrAssume}) to act with $\bar S_{1\to1}$ on $\bS^{(2)}[J]$
\beqa\la{ordertwo}
\bar S_{1\to1}\bS^{(2)}[J]&=&-g{\cL\over 2}\(\bar S_{1\to2}\widetilde \bS^{(1)}[J]+\bar S_{2\to1}\widetilde \bS^{(0)}[J]\)\\&=&-g\bar S_{1\to2}\widetilde \bS^{(2)}[J]-g\bar S_{2\to1}\widetilde \bS^{(1)}[J]- g \bar S_{3\to 0} \widetilde \bS^{(0)}[J]~.\nn
\eeqa
The new term appearing in (\ref{ordertwo}) is (see figure \ref{S30fig})
\beqa\la{Sthreezerobare}
\bar S_{3\to 0} &=&{1\over 2}[\cL,\bar S_{2\to1}]={1\over 2}[\cL,[\cL,\bar S_{1\to2}]]\\
&=&{1\over 2}{\sum_{s,s_1,s_2=\pm}}\!\!\!\!\!'~ss_1s_2\int dy_1dy_2\,G_{12}\int d^{4|4}\Lambda d^4\eta'd\alpha\bar\lambda\eta'\Tr\[\hat{\check J}(\widetilde\Lambda^s_{y_1+y_2})\check J(-\widetilde\Lambda^{s_2}_{2,y_2})\check J(-\widetilde\Lambda^{s_1}_{1,y_1})\]~,\nn
\eeqa
where
$$
G_{12}=-{1\over 3}\[{1\over y_1+is_10}{1\over y_2+is_20}-{1\over y_1+is_10}{1\over y_1+y_2+is0}-{1\over y_1+y_2+is\,0}{1\over y_2+is_2\,0}\] ~.\nn
$$
represents the propagators in the three possible combinations represented in figure \ref{S30fig}. Similarly to (\ref{delta})  we now have $G_{12}= \pi ^2\delta(y_1)\delta(y_2)\(1+s_1s_2\)\(1-ss_1\)/3$.
Plugging it back into (\ref{Sthreezerobare}), we conclude that
\beq
\hspace{-2mm}
\boxed{
\bar S_{3\to0}={2\pi^2\over 3}\sum_{s=\pm}s\int d^{4|4}\Lambda d^4\eta'd\alpha\bar\lambda\eta'\Tr\[\check J(\Lambda^s)\check J(-\Lambda^s_2)\check J(-\Lambda^s_1)\]}~. \la{S30fin}
\eeq
\begin{figure}[t]
\epsfxsize=15cm
\centerline{\epsfbox{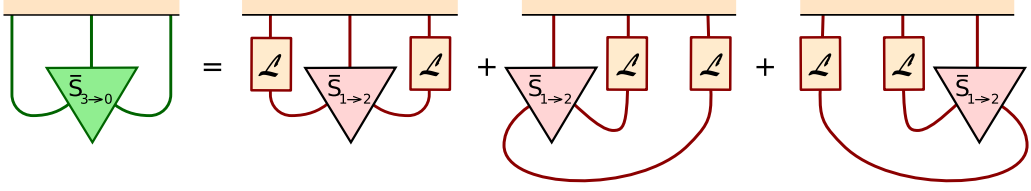}}
\caption{\small\small\textsl{The generator $\bar S_{3\to 0}$ obtained by composing twice the propagator inserting  operator $\mathcal{L}$  with the splitting generator $\bar S_{1\to 2}$. }}\label{S30fig}
\end{figure}
Since $\bar S_{3\to0}$ does not have external legs (to be contracted with $\cL$), it commutes with the propagator inserting operator
$$
[\cL,\bar S_{3\to0}]=0~.
$$
Thus, iff we now go to higher orders in out recursive argument, no new corrections to the special superconformal generator $\bar S$ are produced. Collecting the terms in the $x$ expansion of $\bar S_{1\to1}\bS[J]$ we conclude that
\beq
\(\bar S_{1\to1}+g\bar S_{1\to2}+gx\bar S_{2\to1}+gx^2\bar S_{3\to0}\)\bS[x,J]=0~. \nn
\eeq
By setting $x=1$, we obtain
\beq
\hspace{-2mm}
\boxed{
\(\bar S_{1\to1}+g\bar S_{1\to2}+g\bar S_{2\to1}+g\bar S_{3\to0}\)\bS[J]=0}~. \la{main}
\eeq

A few comments are in order:
\begin{enumerate}
\item[$\bullet$] Note that even though in our derivation we used a generalization of CSW (\ref{CSWloop}) that technically involved a choice of a reference null vector $\chi$, it dropped out of all our final results.
\item[$\bullet$] Naively, the action of $\cL\bar S_{1\to1}$ on a connected amplitude also produces an holomorphic anomaly proportional to $\delta^2(\<l,-l\>)$. Due to the Grassmanian integration, there is no such contribution (see Fig 6.b).
\item[$\bullet$] The operator $\bar S_{2\to1}$ contributes only when the two collinear particles have the same energy sign. The reason for that is a cancelation between the two terms in $[\cL,\bar S_{1\to2}]$ (see figure 7).
In that sense, $\bar S_{2\to1}$ is different from the tree level corrections $S_{1\to2}$ where the two collinear particles can have opposite energy sign \cite{Bargheer:2009qu}. Generalizing these two generators with $\bar S_{3\to0}$, we see that all corrections to the generators are made from the same three vertex connected to the amplitude by cut propagators.
\end{enumerate}

The corrections to the conjugate special superconformal generator $S$ are obtained in an identical way using anti-MHV CSW rules
\beq\la{Scorrections}
\boxed{
\begin{array}{ll}\displaystyle
S_{2\to1}& \displaystyle =-2\pi^2\sum_{s=\pm}\int d^{4|4}\Lambda d^4\eta' d\alpha \,\delta(\eta') \lambda \partial/\partial \eta' \Tr\[J(\Lambda^s)\check J(\Lambda_1^s)\check J(\Lambda_2^s)\]\\ \displaystyle
S_{3\to0}& \displaystyle =-{2\pi^2\over 3}\sum_{s=\pm}\int d^{4|4}\Lambda d^4\eta'd\alpha\delta(\eta') \lambda \partial/\partial \eta'\Tr\[\check J(-\Lambda^{s})\check J(\Lambda^s_2)\check J(\Lambda^s_1)\]
\end{array}}~,
\eeq
where $\Lambda_1$ and $\Lambda_2$ are given in (\ref{Grassman2}). The special conformal generator can be obtained by commuting the superconformal generators as in (\ref{SSK}). This is one of the advantages of the approach of this section. Namely, since we obtain the corrected generators at one loop by acting on the tree level generators with the propagator inserting operator $\cL$, we automatically get the good commutation relations for free: it suffices to conjugate (a straightforward off-shell generalization of) the commutation relations of  \cite{Bargheer:2009qu} by the propagator inserting operator $\cL$.  

We can now read the transformation of the full S-matrix $\bS_{matrix}$ (\ref{Smatrix}) by multiplying (\ref{main}) by $e^F$ (\ref{eqF}) and commuting this through the generators. In this way we obtain\footnote{Similarly to $[\cL,\bar S_{1\to1}]=0$, we have $[F,\bar S_{1\to1}]=0$ because (\ref{Sbarl}) is a total derivative.}
\beq
\hspace{-2mm}
\boxed{
\(\bar S_{1\to1}+g\bar S_{3\to0}+g\bar S_{0\to3} \)\bS_{matrix}[J]=0}~, \la{main2}
\eeq
where
\beq\nn
\boxed{
\bar S_{0\to3} \displaystyle=-{2\pi^2\over 3}\sum_{s=\pm}s\int d^{4|4}\Lambda d^4\eta'd\alpha\bar\lambda\eta'\Tr\[J(\Lambda^s) J(\Lambda^s_2) J(\Lambda^s_1)\]}~.
\eeq
The correction $\bar S_{0\to3}$ do not contain functional derivatives $\check J$ and with this respect it is distinct from $\bar S_{1\to 1}$ and $\bar S_{3\to 0}$. Note also that $\bar S_{2\to 1}$ and $\bar S_{1\to 2}$ are automatically reproduced from commuting $\bar S_{3\to 0}$ through $e^F$ and need not be included in \rf{main2}.

\subsection{Superconformal Invariance of the Tree Level S-matrix}
\begin{figure}[t]
\epsfxsize=16cm
\centerline{\epsfbox{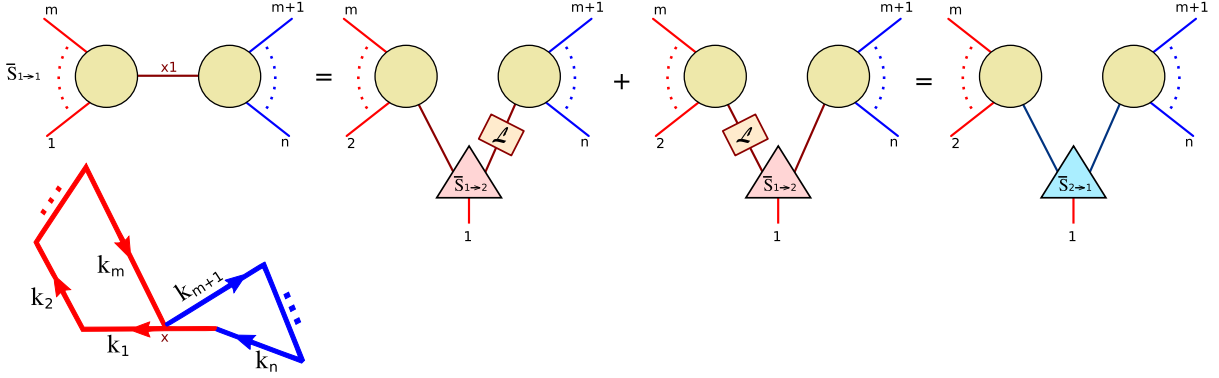}}
\caption{\small\small\textsl{The action of the bare superconformal generator $\bar S_{1\to1}$ on an amplitude at the point where it factorizes on a multi particle pole result in a new anomaly. The anomaly is identified with correction  $\bar S_{2\to1}$ to $\bar S_{1\to1}$, acting on two disconnected amplitudes. In the T-dual polygon Wilson loop picture of \cite{AM}, the collinear multi particle factorization points correspond to configurations where a cusp collides  with an edge. The superconformal generator mixes that configuration with two disjoint polygons touching at a point.} }\label{newTree}
\end{figure}
Our derivation of the main result \rf{main} is valid at any loop order. In this section we will consider the implication of this relation for tree level amplitudes.

The first term is the only one which survives for generic configurations of external momenta and was considered for MHV amplitudes in \cite{Witten} and for all tree level amplitudes in \cite{tree1, dualpapers3}. The second term arises when two of the external momenta become collinear and was proposed in  \cite{Bargheer:2009qu} as the correction to the bare superconformal generator. The last two terms in (\ref{main}) contribute already at tree level and were overlooked in the literature. In this subsection we shall explain for which configuration of external momenta they become relevant.

We shall start by $\bar S_{2\to 1}$. Whenever a subset of adjacent momenta becomes on-shell
$$
(k_i+\dots+k_{i+m-1})^2=P^2=0~,
$$
the amplitude factorize as
\beq\la{Multiparticlefactorization}
\cA_n(k_1,\dots,k_n)\quad\rightarrow\quad-i\int d^4L\int d^4\eta_P\cA(L,k_i,\dots,k_{i+m-1}){1\over L^2+i0}\cA(-L,m+i,\dots,i-1)~.
\eeq
That property of scattering amplitude follows directly from the unitarity of the S-matrix and is called multi particle factorization. The right hand side of (\ref{Multiparticlefactorization}) is nothing but two disconnected amplitudes joined by our propagator inserting operator $\cL$. When acting with $\bar S_{1\to1}$, we can follow exactly the same steps as in the previous subsection (see figure \ref{newTree}). The result is therefore equal to $\bar S_{2\to1}$ acting on two disconnected amplitudes. It is non-zero whenever $P$ become collinear to one of the neighboring momenta $k_i$, $k_{i-1}$, $k_{i+m}$ or $k_{i+m+1}$. The only difference from acting on a connected piece of the amplitude is the absence of additional propagators connecting the two sub-amplitudes. These however, played no role in our previous derivation.

\begin{figure}[t]
\epsfxsize=16cm
\centerline{\epsfbox{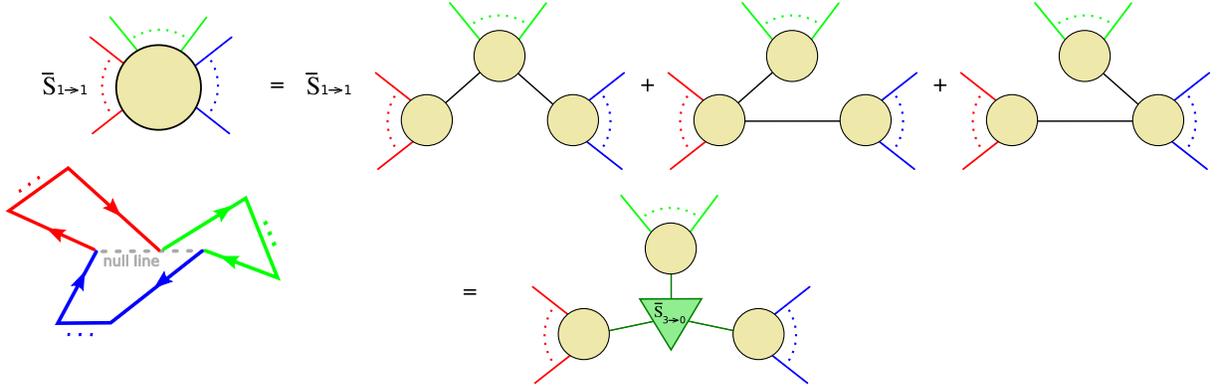}}
\caption{\small\small\textsl{The action of the bare superconfornal generator $\bar S_{1\to1}$ on an tree level   amplitude with a multi-particle pole. The result can be recast as $\bar S_{3\to 0}$ connecting three disconnected tree amplitudes. At one loop level the correction $\bar S_{3\to 0}$  plays a role at single multi-particle pole whereas starting from two loops this corrections becomes generically present (similarly to $\bar S_{2\to 1}$ at one loop). } }\label{newTree2}
\end{figure}

The last contribution, $\bar S_{3\to 0}$, arises at a double multi particle pole
\beq
(k_{i}+\dots+k_{j})^2,(k_{j+1}+\dots+k_{l})^2 \to 0 \,\,\, , \,\,\, i<j<l
\eeq
where the two subset of null momenta also become collinear\footnote{At that point, due to momenta conservation, the remaining momenta $k_l+\dots k_n$ will automatically become null and collinear with the two previous null subset of momenta.}. The derivation of this correction follows precisely as before. When the bare generator acts on an amplitude with such kinematics the relevant CSW diagrams are those in figure \ref{newTree2}.\footnote{This figure represents the relevant part of a bigger CSW graph, i.e. the dots could stand for extra propagators connecting to more MHV vertices} Two collinear legs are connected to one of the  MHV vertices. The holomorphic anomaly associated with these two collinear legs generates a leg splitting operator $\bar S_{1\to 2}$ acting on an MHV vertex with one fewer leg. The two legs coming out of $\bar S_{1\to 2}$ are then connected to the other pieces of the graph. The sum over the three possible diagrams in figure \ref{newTree2} generates the correction $\bar S_{3\to 0}$ by exactly the same mechanism as explained in the main text.

\section{Regularized Generators and Conformal Invariance of the Regularized S-matrix}\la{sec:reg}

In the previous section we have seen that formally, the S-matrix is superconformal invariant. That analysis was formal because $\cN=4$ SYM is conformal and therefore doesn't have asymptotic particles. However, the S-matrix observables we are interested in are IR safe quantities like an inclusive cross section. To compute these and argue for their superconformal covariance, one must first introduce an IR regulator. The IR regulated theory has an S-matrix from which the desired observables are computed. A good IR regulator is a regulator that drops out of IR safe physical quantities leaving a consistent  answer behind.\footnote{The most commonly used regularization is dimensional regularization in which the external particles and helicities are kept four dimensional while the internal momenta are continued to $D=4-2\epsilon$ dimensions (where $\epsilon<0$). It is a good regularization however, it smears the holomorphic anomaly which makes it hard to separate the correction to the generators from a conformal anomaly. Moreover, we have seen that the bare generators mixes internal momenta with external ones. Therefore, the regularized generator $\bar S_{2\to1}$ for example, must act on an amplitude where the external momenta are treated in the same way as the internal ones (and therefore can carry momentum in the $-2\epsilon$ directions).}

\begin{figure}[t]
\epsfxsize=7cm
\centerline{\epsfbox{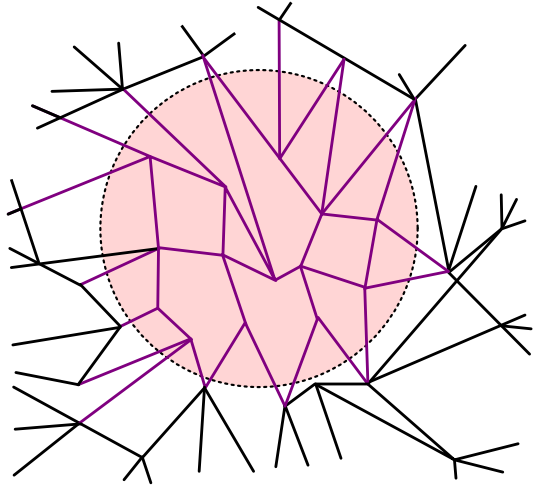}}
\caption{\small\small\textsl{An MHV sub-diagram inside a larger CSW graph. The total off-shellness entering the sub-diagram is zero (\ref{constraint}). For generic off-shell shifts  of the "external" particles this sub-amplitude is finite. } }\label{MHVfig}
\end{figure}

In this section we will introduce an apparently new regularization scheme which we call {\it sub MHV regularization}. The advantages of this regularization scheme are that it is ``holomorphic anomaly friendly" and that the external momenta are treated in the same way as the internal ones.

It is important to keep in mind that IR safe quantities should exhibit the symmetries of the theory -- in the case at hand superconformal symmetry -- however we can have perfectly suitable regulators which break part of the symmetry when considering intermediate regulator dependent quantities\footnote{E.g. lattice regularization typically breaks most of the symmetries of the continuous theory and yet it is the regularization which (almost always) leads to the most reliable and rigorous results.}. This is certainly the case for our proposal:  we suggest a regulator which preserves the superconformal symmetry generated by $\bar S$;  however the  conjugate symmetry generated by $S$ (and therefore also special conformal transformations) will not be a symmetry of the regulated S-matrix. As for Lorentz invariance, the CSW procedure picks up a particular null momenta $\chi$ which can be thought of as a choice of light-cone gauge \cite{axial}. The regularized S-matrix depends on $\chi$ and the Lorentz generators also rotate this vector. Similarly we could design a conjugate regularization which would preserve the symmetry spanned by $S$. Assuming our proposed regularization to be a good regulator, this implies that IR safe (regulator independent) quantities  will be invariant under both $S$ and $\bar S$ (and therefore also $K$).

We saw before that in order to go from trees to loops it was quite useful to generalize scattering amplitudes to their tilded counterparts where external legs are put off-shell by means of the CSW prescription. These off-shell amplitudes appear quite naturally when
we consider a sub-diagram inside a larger CSW expansion, (see figure \ref{MHVfig}). In this sub-diagram each ``external" leg is characterized by a null momenta $l_i$ and an off-shell momenta $L_i=l_i+\delta y_i\ \chi$ such that
\beq
\sum_{j=1}^n \delta y_j =0 \la{constraint}~,
\eeq
where $n$ is the number of ``external" legs. For generic values of the of-shell shifts ($\delta y_j$) the sub-amplitudes obtained in this way are finite. This leads us to suggest a
\textit{sub MHV regularization} of scattering amplitudes: we replace any external (on-shell) momenta $l_i$ by an off-shell momenta of a small mass $m_i$ such that\footnote{The mass $m_i$ is taken small with respect to the amplitude Lorentz invariants.}
\beq\la{Reg}
l_i\quad\longrightarrow\quad L_i=l_i+\delta y_i\ \chi~,\qquad\delta y_i={m_i^2\over 2l_i\cdot\chi}~,
\eeq
where $\chi$ is an arbitrary null momenta and (\ref{constraint}) is imposed.
The regulated amplitude is then given by the corresponding CSW sum where the external off-shell momenta are treated in the same footing as internal ones.
Our proposal for the regulated S-matrix is just the same as considered before (\ref{CSWloop}). The only difference is that now we use it to generate slightly off-shell (\ref{Reg}) processes. That is, as internal $J$'s, all external $J$'s are functions of the on-shell momenta and superspace coordinate $\Lambda_i$ and the of-shellness parameter $y_i$.  Notice that (\ref{constraint}) automatically leaves the total momenta undeformed.
Furthermore, it can be implemented at the level of the tree level functional $\exp \widetilde\cA^{(0)\rm MHV}_c[J,\chi] $ because the propagator inserting operator $\mathcal{L}$ always acts on legs with opposite $y$'s and thus does not spoil this condition.\footnote{We expect that up to sub-leading terms in the regulator, this regularization is equivalent to the more conventional regularization in which the external particles are given a small mass $m_i$.}

It is instructive to understand what are the sources of divergence in one loop MHV amplitude (\ref{oneloopcut}) and why these are regulated by (\ref{Reg}). For generic value of $y$, and after dressing out the tree level amplitude, the dLIPS integral in (\ref{oneloopcut}) computes the discontinuity of a one-loop amplitude in a multi-particle channel. The result is finite, unless $P_{L,y}$ is equal to a linear combination of two momenta adjacent to the cut (these are $k_i$, $k_{i+m-1}$, $k_{i+m}$ and $k_{i-1}$). Near such a point $y_\star$, the dLIPS integral behaves as $\log(y-y_\star)$. It will lead to a divergence of the $y$-integral only if $y_\star=0$, where the $\log$ singularity coincide with the simple pole in the measure. That is the case for $m=1$ and $m=2$ only \footnote{Other than that, the simple pole at $y=0$ (with an $i0$ prescription) can lead to a divergence only if $y=0$ is a limit of integration. That is the case for $m=1$ only. We conclude that before regularization, the only divergent integrals in the sum (\ref{oneloopcut}) are the ones with $m=1$ and $m=2$.}.
In both cases the divergences come from the region of the $y$-integration near $y=0$ where the integrand behaves as
$
{\log(y-i0)\over y+i0}\,.
$
For $m=1$, the point $y=0$ is also the limit of integration (this means that the divergences from both sides of the pole do not cancel each other and therefore the $m=1$ term diverges more severely than the $m=2$ contribution). For $m=2$, the point $y=0$ is not the limit of integration however the contour of integration is trapped between the $\log$ and the pole singularities. After regularization (\ref{Reg}), the simple pole and the log singularity are separated by $\delta y_i$ for $m=1$ and by $\delta y_i+\delta y_{i+1}$ for $m=2$. Moreover, for $m=1$ the location of the simple pole differ from the bottom limit of integration by $\delta y_i$. The resulting integrals (with $i0$ prescription) are therefore finite.

 In what follows, we will assume MHV sub regularization to be a good regularization. \footnote{We plan to consider this regularization in greater detail elsewhere.} We will now show that the superconformal generator $\bar S$ can be easily deformed in such a way that it is still a symmetry of the regularized S-matrix.

The derivation of the regulated generators follows exactly the same steps as in the previous formal section. The only difference is that now the external momenta also carry an off-shell component parametrized by $\delta y_i$ (\ref{Reg}). Again, since MHV mass regularization leaves the MHV vertices untouched, so are the holomorphic anomalies. The resulting regularized form of the generators $\bar S_{1\to 1}$ and $\bar S_{2\to 1}$ are almost identical to their on-shell counterparts and read
\beqa\la{Reggen}
(\bar{{S}}_{{1\to 1}})_{Reg}&=&-\sum_{s=\pm}s\int d^{4|4}\Lambda\,dy\, \eta  \,  \Tr\[\bar\partial J(\widetilde\Lambda^s_y) \check{J}(\widetilde\Lambda^s_y)\]  \\
(\bar S_{2\to1})_{Reg}&=& \pi{\sum_{s,s_1,s_2=\pm}}\!\!\!\!\!'~s\int d^{4|4}\Lambda d^4\eta'd\alpha\ dy\ \bar\lambda\eta'\\
&\times&\int dy'\[{i\over y'+y+i0}-{i\over y'-s_1s_2i0}\]  \Tr\[J(\widetilde\Lambda^s_y)\check J(\widetilde\Lambda^{s_1}_{1,y'+y})\check J(\widetilde\Lambda^{s_2}_{2,-y'})\]~. \nn
\eeqa
When acting with $(\bar S_{2\to1})_{Reg}$ on the regularized amplitude a new loop is formed. In that loop, $y$ is the off-shell regulator. For any non zero value of $y$, we don't have the cancelation obtained in the formal discussion around (\ref{intygen}). Instead, $(\bar S_{2\to1})_{Reg}$ is connected to the amplitude by the difference between two off-shell propagators. As we try to remove the regulator, these propagators become more and more concentrated around the on-shell point. The generator $\bar S_{3\to 0}$ does not involve external legs and therefore stands as it is (\ref{S30fin}) after regularization;  the off-shell generator $\bar S_{1\to 2}$ is given in (\ref{offS12}).  We conclude that the formal structure obtained in the previous section survives regularization.

It would be very interesting to repeat our analysis, both formal and regularized, for the dual conformal symmetry. We expect to be able to prove, under the same assumptions in this paper, dual conformal invariance of scattering amplitudes at any loop order. Furthermore, it would be very interesting to understand how these two symmetries combine into a Yangian at loop level. Finally, an obvious question is of course to which extent can we use all these symmetries for computational purposes. We plan to address these issues in a separate publication.

\section*{Acknowledgements}

We would like to thank  N.Beisert, N.Gromov, J.Henn, V.Kazakov, T.McLoughlin, J.Penedones, J.Plefka and D. Skinner for interesting discussions. We would specially like to thank F.Cachazo for many enlightening discussions, suggestions and for referring us to many relevant references. The research of AS has been supported in part by the Province of Ontario through ERA grant ER 06-02-293. PV thanks the Perimeter Institute for warm hospitality during the concluding part of this work.

\appendix
\section{Appendix. The $m=1$ case}\label{AppA}

\begin{figure}[t]
\epsfxsize=7cm
\centerline{\epsfbox{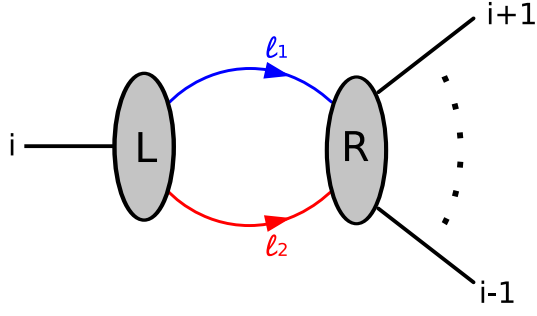}}
\caption{\small\small\textsl{When writing the MHV one loop amplitude as a CSW sum we also have terms where a single external leg $k_i$ stands alone in an MHV 3 vertex. When acting with $\bar S_{1\to 1}$ the holomorphic anomalies might set one of the internal momenta to be collinear with $k_i$. Cutting the internal leg in this way would generate bivalent MHV vertices. In appendix \ref{AppA} we explain that these contributions actually vanish.}}\label{appAfig}
\end{figure}
The MHV one loop amplitude when written as a sum of dispersion integrals contains some peculiar terms -- depicted in figure \ref{appAfig} -- where a single external particle on the left is connected by a 3-MHV vertex to the remaining $n-1$ external particles in a tree level amplitude $\mathcal{A}_{n+1}$. When acting with $\bar S_{1\to 1}$ on these contributions we will again set the internal lines $l_1$ or $l_2$ to be collinear to one of the external legs: either the leg $i$ on the left or one of the legs $i-1$ or $i+1$ on the right. The latter case poses no problem: e.g. when we cut the line $l_1$ making it collinear with $i+1$ we obtain a three vertex on the left connected with a $A_{n}$ amplitude on the right; this case is covered in the main text. Much more problematic at first sight is the former contribution: if $l_1$ becomes collinear to $i$ then naively we will obtain a bivalent MHV vertex on the left connected to a $A_{n+1}$. Bivalent vertices are of course not present in the CSW rules and this could cause a problem.\footnote{See also section \textit{A subtle detail} in \cite{CSW}} In this appendix we will consider these contributions and explain that they actually vanish.

Without loss of generality let us take $i=1$. We need to collect from  \beqa
\cA_n^{(0)}\int{dy\over y+i0}\int d^4l_1d^4l_2\delta^{(+)}(l_1^2)\delta^{(+)}(l_2^2)\delta^4(k_1-(y-y')\chi-l_1-l_2)\int d^4\eta_{l_1}d^4\eta_{l_2}\delta^8(Q_R-Q_l) \nn \\
 \(\bar S_L+\bar S_R+\bar S_{l_1,l_2}\){\<n1\>\<12\>\over\<1l_1\>\<2l_1\>\<1l_2\>\<nl_2\>\<l_1l_2\>^2} \nn
\eeqa
the terms where an internal momentum becomes collinear to $k_1$. Recall that in our regularization each leg has some $y_j$ which measures the particle off-shellness; $y_1\equiv y'$. Preforming the Grassmanian integrations we find the following expression for those terms:
\beqa
\!\!\!\!\!\!\!\!\!\!&&-{4\pi^2 ic(\epsilon)\over\sin(\pi\epsilon)}\cA_n^{(0)}\eta_1\!\!\int\!{dy\over y+i0}\!\int \!d^4l_1d^4l_2\delta^{(+)}(l_1^2)\delta^{(+)}(l_2^2)\delta^4(k_1\!-\!(y-y')\chi\!-\!l_1\!-\!l_2){\<n1\>\<12\>\<l_1l_2\>^2\over \<2l_1\> \<nl_2\>}  \nn\\
\!\!\!\!\!\!\!\!\!\!&&\times\(\frac{\<l_1l_2\>\bar\lambda_{l_1}-\<1l_2\>\bar\lambda_1}{\<1 l_2\>\<l_1 l_2\> } \delta^2(\<1l_1\>)+\frac{\<l_1l_2\>\bar\lambda_{l_2}-\<l_11\>\bar\lambda_1}{\<1l_1\>\<l_1 l_2\>} \delta^2(\<1l_2\>)+2 \frac{\<1l_2\>\bar\lambda_{l_2}-\<l_11\>\bar\lambda_{l_1}}{\<1l_1\> \<1 l_2\> }\delta^2(\<l_1l_2\>)\). \la{second}
\eeqa
We will now see that each of the three terms inside the parentheses leads to a vanishing expression by itself. We start by considering the first term. Preforming the integration over $l_2$ using the momentum conservation delta function we see that this term is proportional to
\beqa
\int\!{dy\over y+i0}\!\int \!d^4l_1 \delta^{(+)}(l_1^2)\delta^{(+)}(l_2^2) { \<l_1l_2\>\over \<2l_1\> \<nl_2\>\<1 l_2\>}  \(\<l_1l_2\>\bar\lambda_{l_1}-\<1l_2\>\bar\lambda_1\) \delta^2(\<1l_1\>)\,.
\eeqa
where $l_2=k_1-(y-y') \chi-l_1$. Next we use the delta function $\delta^2(\<1l_1\>)$ to set $l_1=t k_1$. The previous integral is then proportional to
\beqa
\int \frac{dy}{y+y'} \int dt  (1-t) { y\<1|\chi|\rho]\over (1-t)\<n|k_1|\rho]-y\<n|\chi|\rho]} \delta(y(1-t))=0  \la{B1} \,.
\eeqa
Here $\rho$ is an arbitrary spinor used to simplify
\beqa
\frac{\<1 l_2\>}{\<n l_2\>}=\frac{\<1 l_2\>[l_2 \rho]}{\<n l_2\>[l_2 \rho]}={- y \<1|\chi|\rho]\over (1-t)\<n|k_1|\rho]-y\<n|\chi|\rho]} \,
\eeqa
and changed variables from $y-y'\to y$.
Notice that without the $y'$ regularization the integral (\ref{B1}) would be ill defined as mentioned in the main text. The second term in (\ref{second}) is treated similarly. This leaves us with the last term, proportional to $\delta^2(\<l_1l_2\>)$. We first use (twice) the Schouten identity to re-write the spinor ratio pre-factor as:
$$
C\equiv {\<12\>\<1n\>\<l_1l_2\>^2\over\<1l_1\>\<2l_1\>\<1l_2\>\<nl_2\>}=-\frac{\left\langle 2 l_2\right\rangle  \left\langle n l_1\right\rangle
   }{\left\langle 2 l_1\right\rangle  \left\langle n l_2\right\rangle
   }+\frac{\left\langle 1 l_2\right\rangle  \left\langle
   n l_1\right\rangle }{\left\langle 1 l_1\right\rangle  \left\langle
   n l_2\right\rangle }+\frac{\left\langle 1 l_1\right\rangle
   \left\langle 2 l_2\right\rangle }{\left\langle 1 l_2\right\rangle
   \left\langle 2 l_1\right\rangle }-1$$
Notice that for $l_1 \propto l_2$ each term is either $1$ or $-1$ and the sum of all terms is zero. With the off-shell regularization the remaining terms do not lead to divergencies and therefore this contribution also vanishes.

\printindex

\end{document}